\documentclass[apj]{emulateapj}
\usepackage{epstopdf}
\def\gsim{\ \raise 3pt \hbox{$\rangle$} \kern -8.5pt \raise -2pt \hbox{$\sim$}\ }
\usepackage{graphicx,natbib,color,amsmath,subfigure}
\newcommand{\blank}[1]{}
\usepackage{color}
\def\rhessi{{\textit{RHESSI}}}
\def\kw{{Konus-\textit{Wind}}}
\def\mw{{microwave}}
\def\Mw{{Microwave}}
\def\EM{\textit{EM}}

\shorttitle{Cold Flare}
\shortauthors{Fleishman et al.}

\begin{document}


%\end{landscape}

\title{A COLD FLARE WITH DELAYED HEATING}

%\author{S. Djorgovski\altaffilmark{1,2,3} and Ivan R. King\altaffilmark{1}}
%\affil{Astronomy Department, University of California,
%    Berkeley, CA 94720}
%
%\author{C. D. Biemesderfer\altaffilmark{4,5}}
%\affil{National Optical Astronomy Observatories, Tucson, AZ 85719}
%\email{aastex-help@aas.org}
%
%\and
%
%\author{R. J. Hanisch\altaffilmark{5}}
%\affil{Space Telescope Science Institute, Baltimore, MD 21218}
%
%%% Notice that each of these authors has alternate affiliations, which
%%% are identified by the \altaffilmark after each name.  Specify alternate
%%% affiliation information with \altaffiltext, with one command per each
%%% affiliation.
%
%\altaffiltext{1}{Visiting Astronomer, Cerro Tololo Inter-American Observatory.
%CTIO is operated by AURA, Inc.\ under contract to the National Science
%\altaffiltext{2}{Society of Fellows, Harvard University.}
%\altaffiltext{3}{present address: Center for Astrophysics,
%    60 Garden Street, Cambridge, MA 02138}
%\altaffiltext{4}{Visiting Programmer, Space Telescope Science Institute}
%\altaffiltext{5}{Patron, Alonso's Bar and Grill}

\author{Gregory D. Fleishman\altaffilmark{1,2}, Valentin D. Pal'shin\altaffilmark{2}, Natalia Meshalkina\altaffilmark{3}, Alexandra L. Lysenko\altaffilmark{2}, Larisa K. Kashapova\altaffilmark{3}, Alexander T. Altyntsev\altaffilmark{3}}
\altaffiltext{1}{Center For Solar-Terrestrial Research, New Jersey Institute of Technology, Newark, NJ 07102}
\altaffiltext{2}{Ioffe Institute, St. Petersburg 194021, Russia}
\altaffiltext{3}{ISZF, Irkutsk, Russia}


\begin{abstract}
Recently, a number of peculiar flares  have been reported, which demonstrate significant non-thermal particle signatures with a low, if any,  thermal emission, that implies close association of the observed emission with the primary energy release/electron acceleration region.
This paper presents a flare that appears a ``cold" one at the impulsive phase, while displaying a delayed heating later on.
Using HXR data from \kw, microwave observations by SSRT, RSTN,  NoRH and NoRP,  context observations, and 3D modeling, we study the energy release, particle acceleration and transport, and the relationships between the nonthermal and thermal signatures. The flaring process is found to involve interaction between a small and a big loop and the accelerated particles divided in roughly equal numbers between them. Precipitation of the electrons from the small loop produced only weak thermal response because the loop volume was small, while the electrons trapped in the big loop lost most of their energy in the coronal part of the loop, which resulted in the coronal plasma heating but no or only weak chromospheric evaporation, and thus unusually weak soft X-ray emission. Energy losses of fast electrons in the big tenuous loop were  slow resulting in the observed delay of the  plasma heating. We determined that the impulsively accelerated electron population had a beamed angular distribution in the direction of electric force along the magnetic field of the small loop. The accelerated particle transport in big loop was primarily mediated by turbulent waves like in the other reported cold flares.

\end{abstract}


\keywords{acceleration of particles---diffusion---magnetic fields---Sun: flares---Sun: radio radiation---turbulence}



\section{Introduction}

A close casual relationship between the nonthermal particles accelerated in flares due to release of the excessive magnetic energy and plasma heating has come to be known as the Neupert effect \citep{1968ApJ...153L..59N}. Specifically, \citet{1968ApJ...153L..59N} discovered that the soft X-ray (SXR) light curves in a number of flares at the rise phase and up to the SXR peak were well correlated with the running time integral of the impulsive microwave emission from the flare. Currently, the Neupert effect is commonly referred to a similar relationship between the impulsive hard X-ray (HXR) and thermal SXR emissions. One way or the other, the Neupert effect suggests that (at least in some flares) the particle acceleration takes place first giving rise to nonthermal microwave and HXR emissions and then the energy and momentum losses of these accelerated particles result in the thermal response in the form of coronal plasma heating and/or chromospheric evaporations; the heated coronal plasma is then cooling down relatively slowly due to conductive and radiative losses.

It was long ago established  \citep[e.g.,][]{1988SoPh..118...49D} that the impulsive flares demonstrating a clear Neupert effect represent only a fraction of all events. Most recently, with the spectrally resolved X-ray data obtained with Reuven Ramaty High Energy Solar Spectroscopic Imager  \citep[\rhessi,][]{2002SoPh..210....3L} many more low-energy gradual (presumably, mostly thermal) events have been detected. For example, \citet{2008AdSpR..41..988S} %Su et al. 2008
find that about 2/3 of all events are gradual, up to 10\% are impulsive, and up to 20\% are early impulsive flares \citep{2006ApJ...645L.157S, 2007ApJ...670..862S}; only a fraction of the latter class events demonstrates a clear Nuepert effect, which suggests that the relationships between nonthermal and thermal energies are much more complex in the general case \citep[e.g.,][]{2005ApJ...621..482V} than the simple loss-to-heating correspondence implied by the standard Neupert effect.

It has recently been recognized that {some} early impulsive flares are in fact `cold flares' \citep{Bastian_etal_2007, Fl_etal_2011, 2013PASJ...65S...1M} % Masuda et al. 2013
in which no or {so} modest thermal plasma response is detected {that these events are not listed as GOES flares}. The {three} reported cold flares, although all similar in a lack of the thermal response, are, however, noticeably different from each other in a number of other respects. For example, the 30-Jul-2002 cold flare reported by \citet{Fl_etal_2011} is a `tenuous' flare with the thermal number density not exceeding $2\times10^9$~cm$^{-3}$ at the coronal part of the flaring loop. In such cases the plasma heating due to fast electron collisions with the coronal thermal particles is small because the collisions are rare in the tenuous plasma, while the chromospheric evaporation is suppressed by some reason. On the contrary, two other cold flares reported by \citet{Bastian_etal_2007} and  \citet{2013PASJ...65S...1M} were  dense with the thermal number density in excess of $10^{11}$~cm$^{-3}$. In such cases the fast particle losses in the coronal part of the loop are large and the increase of the thermal energy is relatively strong; but, because of high density, the net temperature increase {above the coronal preflare level} is rather modest. %{, $\lesssim 5-7$~MK, which can be taken as a flag to categorize an event as a `cold flare.' }

\begin{figure}\centering
%C:\MyProjects\ColdFlare10mar2002\overview_20020310.pro
\includegraphics[width=0.9\columnwidth]{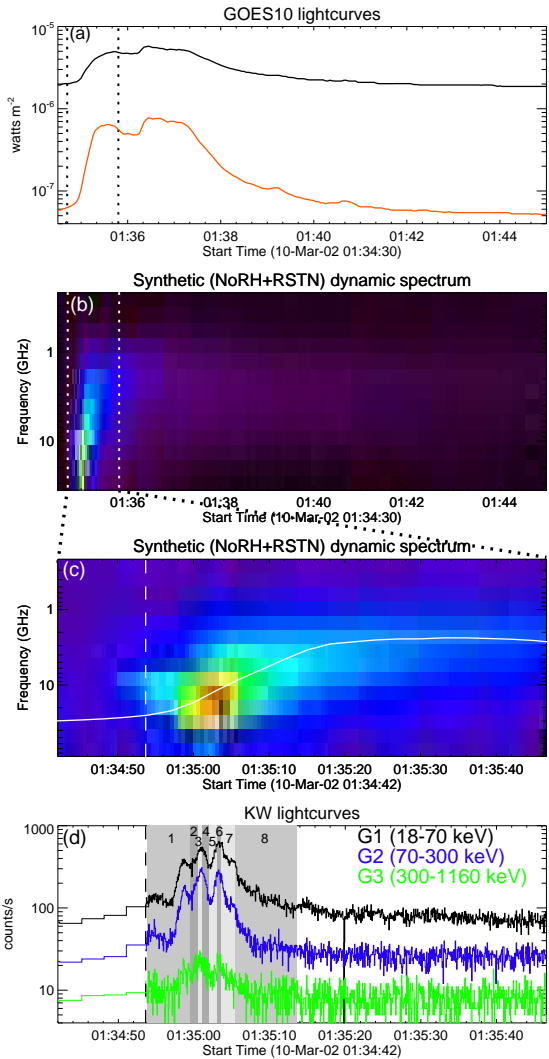}\\
\caption{\label{fig_over_20020310} Overview of  March 10, 2002 flare. (a)  GOES (3\hspace{0.1cm}s) lightcurves
as measured by GOES-10 spacecraft. (b) %assuming photospheric abundances from CHIANTI v5.2 (middle panels).
\mw\ dynamic spectrum. (c) zoom-in dynamic spectrum of the impulsive flare phase. White solid curve shows the high-energy GOES lightcurve.
 (d) \kw\ lightcurves in three energy bands. Dotted vertical lines in (a,b) demarcate the impulsive phase shown in panels (c) and (d). Dashed vertical lines in (c,d) show the start time of \kw\ fast record. The vertical stripes shown by dark or light grey background
denote the eight intervals over which we extracted
the spectra in Figure~\ref{fig_KW_spectra}. %$t_0$(KW)=01:34:53.874~UT.
%(5693.874 s).
 }
\end{figure}

The Neupert effect is clearly present {in the reported cold flares} in its nominal form, i.e., the time derivative of the SXR light curve closely correlates with the light curves of the nonthermal HXR and microwave emissions. Morphologically, the microwave emission in all reported cold flares has a coronal (loop-like) structure, while X-ray morphology differs depending on if the flare is dense or tenuous: the dense flares demonstrate a coronal X-ray source, while the tenuous flare---two chromospheric foot points; the coronal part of the cold flare sources has been identified with the very acceleration region of fast electrons \citep{Fl_etal_2011}. Timing of these events is inconsistent with purely collisional transport of fast electrons , while requires the transport mediated by wave turbulence \citep{Bastian_etal_2007, Fl_etal_2011}.  The spectra of fast electrons are typically hard in the cold flares, with the spectral index $\delta\sim3.5$. Overall, the cold flares appear to be events with efficient electron acceleration but only a modest plasma heating. It appears that the heating is entirely supplied by energy losses of the accelerated electrons without any apparent additional heating. {Therefore, compared with other flare types, the nonthermal-energy-dominated cold flares offer a cleaner way of studying electron acceleration in flares and their effect on subsequent plasma heating. Thus, both the acceleration of electrons and the nonthermal-to-thermal energy evolution can be studied much more conclusively in the cold flares then in a `normal' flare.}
\blank{More cold flares are clearly needed to be analyzed to better understand the processes of particle acceleration, plasma heating, and energy partitions in solar flares.}

This paper presents \blank{one more} {an unusual} case of a solar flare, which shows only a very mild thermal emission during the entire impulsive phase {like other cold flares}, but then demonstrates a more substantial heating {that lasts considerably longer than the impulsive phase} with the heating rate comparable to that observed at the impulsive phase. We argue that this behavior can be understood if two interacting loops are involved in the flare---a small one and a large one. The small loop, presumably the region of its interaction with the big loop, plays a role of particle accelerator and particle injector for the larger loop, where the fast electrons are effectively accumulated. During the impulsive phase the flare thermal response is driven by the fast electron losses from the small loop, but later---by the same losses of the trapped population from the large loop.    %\blank{This implies that the release of the excessive magnetic energy changed the mode during the event: originally, most of the free magnetic energy was converted to the particle acceleration (at the impulsive phase), while then the magnetic energy started to heat the coronal plasma somehow directly, without the step of the electron acceleration, which cannot be reconciled with the standard Neupert effect.}
From the X-ray and microwave data augmented by 3D modeling, we estimate the key source parameters{, such as the total number of the nonthermal electrons, the spectral shape, and even the pitch-angle distributions at a few time frames},  and discuss the corresponding implications for the particle acceleration, plasma heating, and thermal-to-nonthermal energy partitions.

\section{Observations}

\subsection{Data set overview}
\label{S_data_overview}

The solar flare, GOES class C5.1, occurred on 10 March 2002 nearby the eastern solar limb in AR~09866; see summary of the total power data in  Figure~\ref{fig_over_20020310}. HXR and gamma-ray data for this event are obtained with the \kw\ (KW) spectrometer \citep{Aptekar1995}, while the Reuen Ramaty High Energy Solar Spectroscopic Imager  \citep[\rhessi,][]{2002SoPh..210....3L}  data were not available because of the \rhessi\ night. \blank{ The \kw\ has a high sensitivity and high temporal resolution but does not have spatial resolution.}

{\kw\ is a joint US-Russian experiment launched on November, 1, 1994 to study the gamma-ray bursts and solar flares. It consists of two NaI(Tl) detectors S1 and S2 observing correspondingly the northern and  southern celestial hemispheres. {Unlike \rhessi,} this instrument operates in the interplanetary space (since 2004---near Lagrange point L1), so it {does not suffer from} "nights"{, and, thus, has a very high duty cycle of about 95\%}. Thanks to being far from the Earth's magnetosphere  it has an exceptionally stable background. \kw\ works in two modes: waiting mode and trigger mode. In the waiting mode the count rate light curves are available in three {wide} energy channels G1($\sim18$-70~keV), G2($\sim70$-300~keV), G3($\sim300$-1160~keV) with accumulation time 2.944~s. In the trigger mode \kw\ measures count rate light curves in the same three channels with {a varying} time resolution from 2 to 256~ms and with total duration of 250~s.  \blank{and} {While in the trigger mode,} 64 multichannel spectra {are taken in addition to the light curves as follows. The} multichannel spectra are measured in two partially overlapping energy bands: $\sim20$-1150~keV and $\sim240$~keV-15~MeV in 2002 year. Each band has 63 {energy} channels {with fixed nominal boundaries}. Accumulation time of the first four spectra is {fixed at} 64~ms and of the last eight spectra---at 8.192 s.  For \blank{other} {the remaining} 52 spectra the accumulation time {is adaptively adjusted} \blank{varies} from 0.256 s to 8.192 s {based on} \blank{according to} the count rate in the G2 channel: for more intense events the accumulation time is \blank{smaller} {proportionally shorter}. Switch on to the trigger mode occurs at a statistically significant excess above a background count rate within an interval of 1 s or 140 ms in the G2 energy channel} \citep{Palshin_etal_2014}.

The flare triggered the \kw\ detector S2 at $t_0$(KW) = 5693.874 s UT (01:34:53.874). The
propagation delay from WIND to the center of the Earth is 0.241 s for this flare\footnote{The corresponding delay to the Nobeyama Observatory whose microwave data are used for the analysis is 0.231~s.}; % {correcting for this factor, the KW trigger
time corresponds to the Earth-crossing time %5694.115~s UT
01:34:54.115~UT.

Microwave total power (TP) data are obtained with the Nobeyama RadioPolarimeters \citep[NoRP,][]{Torii_etal_1979} in intensity and circular polarization at six frequencies (1, 2, 3.75, 9.4, 17, \& 35~GHz) and in the intensity only at 80~GHz with the time resolution 0.1~s during the impulsive peak, 01:34:36--01:35:46~UT, and 1~s outside the peak (accordingly, no 80~GHz data), as well as 1~s intensity data from the Radio Solar Telescope Network  \citep[RSTN;][]{1981BAAS...13Q.553G} at seven frequencies (0.4, 0.6, 1.4, 2.7, 5.0, 8.8, \& 15.4~GHz)\footnote{To correct for RSTN clock and amplitude calibration errors, the RSTN time was shifted as $t_{\rm true} = t_{\rm obs}-3.3$~s, the light curve at 8.8~GHz was corrected by the factor of 1.4 and at 15.4~GHz by the factor of 1.55. }. %\blank{, and with Siberian Solar Radio Telescope  in intensity and polarization at 5.7~GHz with 14~ms time resolution}.
The flux at 80 GHz was adjusted using the time-dependent correction coefficient $k_{corr}=[T_{years}/1995.83]^{630}$ \citep[Nakajima (2007) private communication; see also][]{Altyntsev_etal_2008, 2009SoPh..260..135K}, %{[GF, Q: Larisa, I obtained a different shape of the 80GHz light curve; we have to check.]},
while the polarization at 1 \& 2 GHz were corrected for the differing gains in the $I$ and $V$ channels \citep[Shibasaki (2007) private communication; see also][]{Altyntsev_etal_2008}. Using these heterogeneous sources of the \mw\ data we built two complementary dynamic spectra of the \mw\ burst. The first of them straightforwardly combines daily NoRP data with RSTN data, both of which have 1~s time (slow) resolution, to form a synthetic dynamic spectrum in the 0.4--35~GHz range. This dynamic spectrum is, however, insufficient for our analysis for two reasons: during the impulsive peak the emission (1) shows subsecond variations and (2) has a high-frequency spectral peak around 35~GHz. Thus, the fast NoRP record made in the burst mode (0.1~s) containing the 80~GHz data is essential for the analysis\footnote{The original 80~GHz data are very noisy; they were smoothed using 4~s window before inclusion into the dynamic spectrum.}. We employ these high time resolution data in two ways. For the light curve and polarization data analysis we use the full time resolution of 0.1~s. But for the spectral analysis we also created a `fast' dynamic spectrum with 0.5~s resolution, which combines the NoRP 0.1~s data resampled to 0.5~s steps with interpolated (from 1~s to 0.5~s) RSTN data. Adding the RSTN data, although compromises the time resolution, is important for the spectral analysis because the NoRP data alone have too few data points for a meaningful spectral fit.

The microwave imaging is performed with the Siberian Solar Radio Telescope (SSRT) at 5.7~GHz (intensity and polarization) and the Nobeyama RadioHeliograph \citep[NoRH,][]{Nakajima_etal_1994} at 17 GHz (intensity and polarization) and 34~GHz (intensity only).
The SSRT is cross-shaped interferometer and the data recorded by the EW and NS arrays provide two-dimensional images of the solar disk every two-three minutes and one-dimensional images every 0.3~s in the standard mode of the observations \citep{SSRT}. The methods of analysis for one-dimensional solar images have been described by Altyntsev et al. (2003) and Lesovoi and Kardapolova (2003). The receiver system of SSRT contains a spectrum analyzer with 120 MHz frequency coverage using an acousto-optic detector with 250 frequency channels, which correspond to the knife-edge-shaped fan beams for the NS and EW arrays. The frequency channel bandwidth is 0.52 MHz. The response at each frequency corresponds to the emission from a narrow strip on the solar disk whose position and width depend on the observation time, array type, and frequency. The signals from all the channels are recorded simultaneously and generate a one-dimensional distribution of solar radio brightness. During the event under study the width of the beam of SSRT was $18''$ for EW array and $30''$ for NS array.

A limited information about the thermal plasma at the flare region is available from a few  images taken at 195~\AA\ {with  Extreme ultraviolet Imaging Telescope (EIT) onboard Solar \& Heliospheric Observatory   \citep[SoHO/EIT,][]{SoHO}}  with the 12 min cadence. The context SXR GOES-10 data and the line-of-sight magnetogram from the Michelson Doppler Imager  \citep[SoHO/MDI][]{SoHO}  are utilized. The line-of-sight magnetogram is used %\blank{later}
for the 3D modeling with the GX Simulator \citep{Nita_etal_2015}.
 %[GF, Q: is there other coronal diagnostics (Trace, SoHO) available for the event?]}

\subsection{Light curves}
\label{S_light_curves}

\begin{figure} %\centering
%C:\MyProjects\ColdFlare10mar2002\cold_flare_20020310_R2.pro :: f2_20020310_R2.ps
\includegraphics[width=0.95\columnwidth,clip]{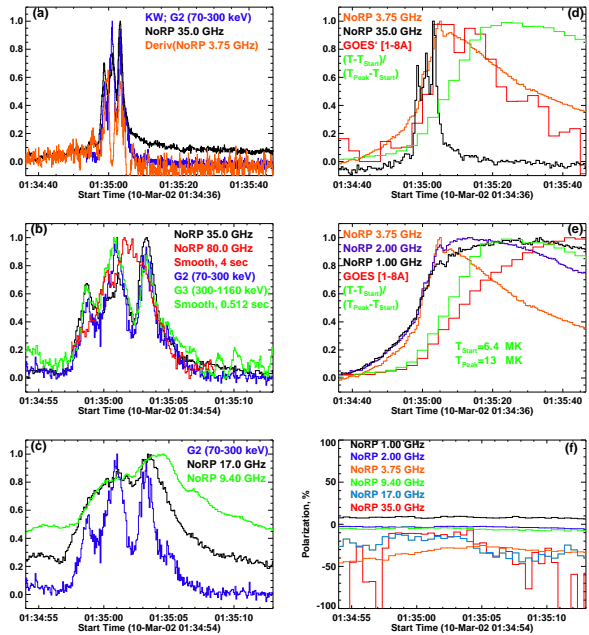}
\caption{Summary of the normalized light curves. (a) Overall comparison of the impulsive \mw\ and HXR light curves at 35~GHz (0.1~s cadence) and 70--300~keV (64~ms cadence) with  the time derivative of the gradual \mw\ light curve at 3.75~GHz. (b)  \Mw\ light curves at 35 \& 80~GHz along with \kw\ HXR G2 and G3 light curves. (c) \Mw\ light curves at 9.4 \& 17~GHz in comparison with \kw\ HXR G2 light curve. (d) \Mw\ light curves at 3.75 \& 35~GHz along with time derivative of the GOES(1--8~\AA) light curve and GOES-derived plasma temperature evolution ($T_{\rm Start}\approx6.4$~MK; $T_{\rm Peak}\approx13$~MK). (e) \Mw\ light curves at 1, 2, \& 3.75~GHz along with GOES(1--8~\AA) light curve and GOES-derived plasma temperature. (f) The degree of polarization of the \mw\ emission at the impulsive flare phase. Note the increase of the degree of polarization at 17 \& 35 ~GHz at the decay phase.
\label{fig_LC} }
\end{figure}

A striking feature of this flare is a contrasting combination of the impulsive and gradual light curves is vividly illustrated in Figure~\ref{fig_LC}:  a prominent impulsive emission is apparent in the KW light curves and in the high-frequency microwave light curve at 35~GHz, while the microwave light curves are getting progressively more gradual towards the lower frequencies; the GOES light curves are even more gradual than the low-frequency microwave light curves. Thus, this section pays a close attention to these various light curves and relationships between them.

Figure~\ref{fig_LC}a shows a general similarity between the \mw\ light curve at 35~GHz, \blank{and} the HXR \kw\ G2 light curve at 70--300~keV, {and time derivative of the \mw\ light curve at 3.75~GHz.} \blank{while Figure~\ref{fig_LC}b offers a comparison between three different \kw\ channels over the impulsive phase of the flare.} Although these three light curves differ from each other in some details, the overall high correlation between them is apparent; no  delay is seen between the light curves.

Figure~\ref{fig_LC}b compares the high energy \kw\ channels G2 and G3 with the {high-frequency \mw} 35~GHz {and 80~GHz} light curves over the impulsive phase of the burst. The \blank{\mw\ } {35~GHz} light curve is as closely correlated with each of the HXR light curves as the HXR light curves correlate with each other. Again, no delay between the \blank{\mw\ } {35~GHz} and HXR light curves is apparent within the instrumental time resolution (64~ms in case of \kw\ and 100~ms in case of NoRP). {On the contrary, the 80~GHz light curve does not correlate in detail with all other light curves, which is the result of the already mentioned 4-s smoothing of the 80~GHz light curve needed to reduce the high fluctuation level in the original signal.} Figure~\ref{fig_LC}c displays a similar relationship, but  between the \kw\ G2 light curve and the \mw\ light curves at 17~GHz and 9.4~GHz. In this case, the \mw\ light curves are less impulsive than and delayed relative to the HXR light curve, while the shapes of the \mw\ light curves at 9.4 \& 17~GHz are closely correlated with each other at the rise phase{, where they appear earlier than the most impulsive light curves}. \blank{We will see later that these behavior originates from the optical thickness effect in these  light curves.}

\blank{Figure~\ref{fig_LC}d confronts the impulsive and gradual light curves.} Figure~\ref{fig_LC}d compares, at the first place, the impulsive 35~GHz light curve and the derivative of the GOES($1-8$~\AA) light curve. In contrast to the expectation based on the standard Neupert effect, these two light curves do not correlate to each other: even though the GOES derivative does reach the peak value at the impulsive peak, it appears strongly delayed relative to the impulsive light curve.
For further reference, this panel also displays the GOES-derived evolution of the plasma temperature and a more gradual \mw\ light curve at 3.75~GHz, which shows a much closer correlation with the GOES derivative. Then, Figure~\ref{fig_LC}e displays all the gradual low-frequency \mw\ light curves, the GOES($1-8$~\AA) light curve, and the temperature evolution. This comparison shows that the light curves are getting more and more gradual and delayed at lower frequencies, with the GOES light curve being the most delayed. {This delay, even though appears small (about 40~s) by the absolute value, is highly substantial indicating the heating process that is roughly four times longer than the duration of the impulsive phase.} The temperature light curve is well correlated with the 1~GHz light curve at their peak phase. It is interesting to recall here, Figure~\ref{fig_LC}a, that the impulsive 35~GHz and HXR light curves are well correlated with the time derivative of the 3.75~GHz light curve (rather than SXR light curve). This correlation indicates that the 3.75~GHz light curve either represents the plasma thermal response on the accelerated electron impact or corresponds to a trapped population of fast electrons, whose injection profile corresponds to the 35~GHz or \kw\ light curves.

Figure~\ref{fig_LC}f shows evolution of the degree of polarization of the \blank{low-frequency} \mw\ emission during \blank{the entire burst (a) and that supplemented with the high-frequency emission during} the impulsive peak. Two interesting features are to be noted about this figure: (i) an unexpectedly high degree of polarization at 3.75~GHz during the entire event\footnote{A comparably strong polarization (not shown in the figure) is detected at 5.7~GHz with SSRT.}{, indicating an optically thin emission at these intermediate frequencies,} and (ii) the degree of polarization at 17 \& 35~GHz is getting larger at the decay of the impulsive phase{, which may imply substantial modification of the angular distribution of the nonthermal electrons; we return to these points later}.

\subsection{Spectra}

\subsubsection{X-ray Spectra}
\label{s_kw_spec_fit}

{\blank{During 445.696~s after the \kw\ trigger time, %\kw\
it measured 64 spectra in two partially overlapping energy bands (from 20 keV to 14 MeV) with accumulations times varied from 64 ms near the trigger time to 8.192~s by the time the signal became undetectable.}}
We performed analysis of the \kw\ spectral data for eight
time intervals  {indicated} in Figure~\ref{fig_over_20020310}d, {where the signal} {\blank{Signal in subsequent time intervals does not} exceed{ed the} background.}  {For the analysis the energy channels}
were rebinned to  {contain} at least 20 counts per energy bin {in each time interval} and fitted in various energy  {subranges} within the 20--1000~keV range, as detailed below. Despite the emission is seen up to $\sim10$ MeV, we did not include the channels above 1~MeV
since they can contain a significant contribution of nuclear line emission, which is not discussed here.

\begin{figure}
%c:\MyProjects\ColdFlare10mar2002\cold_flare_20020310_Xspectrum.pro
%
%\includegraphics[angle=-90, width=0.47\textwidth]{KW20020310_T05693_1_sp7_gr20_BThickEh.eps}
%\includegraphics[angle=-90, width=0.47\textwidth]{KW20020310_T05693_1_sp10_bknPL_style.ps}
\includegraphics[width=\columnwidth]{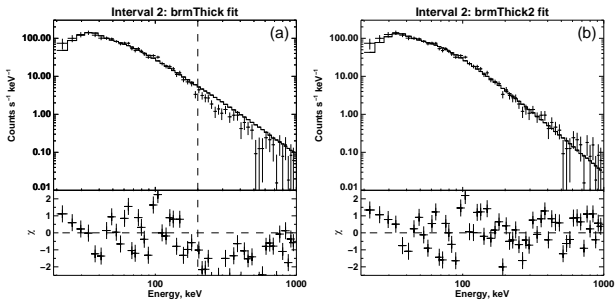}
%\hfill
%\includegraphics[angle=-90, width=0.47\textwidth]{KW20020310_T05693_1_sp10_gr20_BThickEh.eps}\\
%\includegraphics[angle=-90, width=0.47\textwidth]{KW20020310_T05693_1_sp10_brm2PL_style.ps}
%\includegraphics[width=0.47\textwidth]{gSp6BrmThick_GF.eps}
%
%\vfill
%%
%\includegraphics[angle=-90, width=0.47\textwidth]{KW20020310_T05693_1_sp11_gr20_BThickEh.eps}
%\hfill
%\includegraphics[angle=-90, width=0.47\textwidth]{KW20020310_T05693_1_sp12_gr20_BThick.eps}
%%
\caption{\kw\ spectrum of the flare for time interval \#2 (Fig.~\ref{fig_over_20020310}d) with the fit examples: (a) brmThick model fitted to the photon energy range 20--200~keV demarcated by the vertical dashed line, (b) brmThick2 model fitted to the full range of 20--1000~keV.
%{\gff See on-line figure set for other time intervals.}
%\textit{Top Left:} Interval 4.
%\textit{Top Right:} Interval 6. \textit{Bottom Left:} Interval 7s
%\textit{Bottom Right:} Interval 8. И сами рисунки и подписи к ним
%будут значительно улучшены, если мы решим давать их в статье - пока
%я просто вставил стандартные рисунки, которые производит XSPEC, и
%указал для каких они интервалов. {GF: I propose to give two plots for Int 6 and 7, while give all other fits as on-line Figure set.}
}
\label{fig_KW_spectra}
\end{figure}

We fit the spectra with a number of alternative spectral models. Unambiguously, a single power-law (either of the electron or photon spectra) is inconsistent with the data, when the whole range of the photon energies, 20--1000~keV, is analyzed. On the other hand, when a broken power-law model is applied for either electron or photon spectrum, the fit parameters are returned with rather large uncertainties, which implies that a wide range of spectral models  is consistent with the data. In particular, a single power-law can fit the data reasonably well, when a truncated photon energy range is selected, e.g., 20--200~keV (Figure~\ref{fig_KW_spectra}a), 40--400~keV, or 100--1000~keV, which may imply that the spectral steepening progresses slowly but steadily with energy.
{A number of instrumental or physical effects are known to yield spectral flattening at low energies, namely, photon pile-up, photospheric albedo, nonuniform ionization of the target, and return current. We checked via modeling that the pile-up plays no role in our case. Other mentioned effects typically play a role at low energies, $E\lesssim50$~keV \citep{2011SSRv..159..107H}, while in our flare the spectral break happens at much higher energy $E\gtrsim100$~keV. Nevertheless, we employed the fit with albedo correction, but this did not improve the goodness of fit.} {Also we \blank{have} superimposed thermal bremsstrahlung model { with temperature $\sim14$~MK and emission measure $\sim2\times10^{48}~cm^{-3}$ as estimated from the GOES peak flux} on our spectra \blank{with temperature ($\sim14$~MK) and emission measure ($\sim2\times10^{48}~cm^{-3}$) obtained from GOES data.} {and found that the thermal} \blank{But} contribution \blank{of this model} {does not exceed}  1~\%  even at the lowest energy channels.} {As a result of our tests we conclude that the nonthermal electron spectrum has a convex shape (the high-energy slope is steeper than the low-energy one).}
Out of the variety of the considered spectral models we present here some results for three models, using XSPEC 12.5 \citep{Arnaud1996}.

The first of them is a phenomenological broken power-law model, \textit{BPL}, taken in the form:
\begin{equation}
I(E) = \begin{cases} A \left(\frac{E}{100keV}\right)^{-\gamma_1} & E \le E_{br, ph} \\
 A E_{br, ph}^{\gamma_2-\gamma_1} \left(\frac{E}{100keV}\right)^{-\gamma_2} & E_{br, ph} < E \end{cases}
\end{equation}
where $\gamma_1$ and $\gamma_2$ are the \textit{PL} \textit{photon} indexes and $A$ is the
normalization at 100 keV in units of photons~cm$^{-2}$~s$^{-1}$~keV$^{-1}$.

The other two are the collisional thick-target models assuming a power-law  (\textit{brmThick}) in the fast electron {flux} spectrum {(el's/keV/s)} over energy between the  low- and high- energy cutoffs:
\begin{equation}
\label{Eq_brm_1}
F(E) = \begin{cases} 0 & E < E_{cut,low} \\
\propto E^{-\delta} & E_{cut,low} \le E \le E_{cut,high} \\
0 & E_{cut,high} < E \end{cases}
\end{equation}
and a broken power-law electron spectrum (\textit{brmThick2}):
%\[
\begin{equation}
\label{Eq_brm_2}
F(E) = \left\{
\begin{array}{ll}
0 & E < E_{cut,low}\\
\propto E^{-\delta_1} & E_{cut,low} \le E \le E_{br,e} \\
\propto E^{-\delta_2} & E_{br,e} \le E \le E_{cut,high} \\
0 & E > E_{cut,high},
\end{array} \right.
\end{equation}
normalized to the total flux of the electrons [electrons/s] above $ E_{cut,low}$.
Since the XSPEC package %did
does
not contain standard models of the thin  or thick targets from a broken power-law distribution of the nonthermal electrons,  routinely used for the analysis of  X-ray spectra of solar flares,  these models  were added by us to
XSPEC based on analogous models used in the OSPEX package \citep{2002SoPh..210..165S}. % (ссылка).
We also performed the corresponding fitting using the OSPEX package from SSW/IDL to cross-check the fitting results and found that the fitting parameters are fully consistent with each other.

Given that the GOES flux was somewhat low during the impulsive flare phase, we examined if the GOES data can constrain the low-energy cut-off in the accelerated electron spectrum. But in fact, no conclusive constraint was obtained, perhaps, because of a mild thermal emission contribution at the GOES range.
Accordingly, the
low-energy cutoff was fixed to 10 keV (i.e. below the \kw\ fitting range).
The mean atomic number of the target plasma, $Z$, was set to 1.2 {to account the contribution from target nuclei heavier than hydrogen}.

\begin{figure}\centering
%'c:\MyProjects\ColdFlare10mar2002\cold_flare_20020310_del_gam_corr.pro
%
%\includegraphics[width=0.47\textwidth]{SF20020310_Delta_HR21.eps}
\includegraphics[width=0.92\columnwidth]{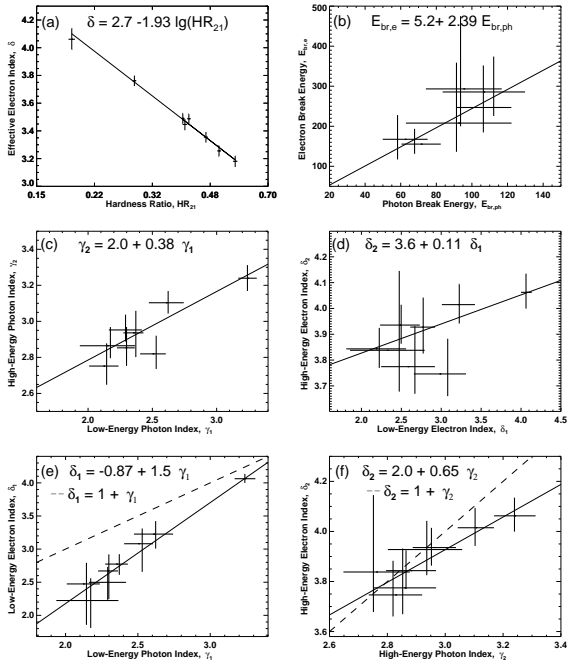}
\caption{Correlation plots for the HXR fit parameters. (a) Cross-correlation between effective power-law index of electron distribution $\delta$, derived from the brmThick fit, Eq.~(\ref{Eq_brm_1}),
and hardness ratio HR$_{21}$ for the eight analyzed spectra; (b) cross-correlation between the break energies in the photon and electron fits; (c) cross-correlation between the low- and high- energy photon indices $\gamma_{1,2}$; (d) cross-correlation between the low- and high- energy electron indices $\delta_{1,2}$; (e) cross-correlation between the photon and electron low-energy indices $\gamma_{1}$ and $\delta_{1}$; (f) cross-correlation between the photon and electron high-energy indices $\gamma_{2}$ and $\delta_{2}$. The dashed lines in (e) and (f) show simple expectation based on classical thick-target model $\delta= \gamma+1$.
%Надо будет доавить еще про аппроксимацию
%{GF: I propose to use black solid line for the regression.}
}
\label{fig_Delta_HR21}
\end{figure}

The fit examples are given in Figure~\ref{fig_KW_spectra}, while the fit  results are summarized in Figure~\ref{fig_Delta_HR21} and in Table~\ref{table_KW_Spectra} (the
errors are given at the 68\% confidence level).
The fact that the HXR spectral analysis could only be performed for a few uneven time intervals {\#1--8 indicated in Fig.~\ref{fig_over_20020310}d} complicates the study of the spectral evolution and comparison with the \mw\ spectral data. However, if the actual electron spectrum does not deviate strongly from a single power-law, the effective spectral slope can be estimated from the hardness ratio, which is the ratio of X-ray fluxes recorded in two adjacent wide energy ranges, G1 and G2, or G2 and G3. To this end we fitted the spectra in our eight available intervals with a single power-law {(\textit{brmThick})} at the photon energy range 20--200~keV and then cross-correlated the obtained spectral index $\delta$ with the hardness ratio $HR_{21}$. Figure~\ref{fig_Delta_HR21}a shows that these two values demonstrate excellent correlation and reveal the following regression law:

\begin{equation}
\label{Eq_hardness2delta}
 \delta = 2.7 - 1.93 \log HR_{21},
\end{equation}
thus, the spectral evolution in this energy range can be recovered with a very high cadence: since the data of the spectral hardness are available with a very high time resolution (down to 16~ms)  the $\delta$ time evolution can be recovered with a comparably high time resolution. %Specifically
We looked but did not find any significant variation of the spectral index $\delta$ on the time scales much shorter than 1~s and; thus, we derived the $\delta$ evolution with 0.5~s cadence needed for comparison with the microwave spectral fit results available with the same cadence.

As has been said, the single power-law does not fit well the data in the entire range 20--1000~keV; therefore, we used a broken power-law over either photon or electron energy. We found the results of the fitting to the photon power-law (BPL) are more stable as compared with the fit to the electron power-law (brmThick2): in the latter case the confidence interval of the fit parameters appears to be rather broad if all the parameters of the broken power-law are kept free, especially, for  intervals 4--7 {(again, we are referring to Fig.~\ref{fig_over_20020310}d)}, while interval  8 can, in fact, be fit by a single power-law model. This behavior of the fit implies that the change of the spectral slope is modest and likely gradual, so no exact value of the break energy could be identified from the data. To get a more stable fit results for intervals 4--7, we first fixed the high-energy slope $\delta_2$ to the values implied by the BPL fits ($\delta_2=\gamma_2+1$) and determined the break energies in the electron spectrum. Then, we cross-correlated the break energies in these two fits, BPL and brmThick2, to determine a regression law between them, Figure~\ref{fig_Delta_HR21}b. At the next step we fixed the break energies at the values implied by this regression law and fit the spectra with free spectral indices. The results of these slightly different versions of the brmThick2 fits agree well to each other, which confirms that the spectral model with a broken power-law is overall consistent with the data. % Although this does not bring any additional information about the spectral shape, these tests .

The remaining panels of Figure~\ref{fig_Delta_HR21} display correlation plots between different pairs of the spectral indices obtained from the fits. %Correlations are obtained using the IDL MPFITEXY routine\footnote{http://user.astro.columbia.edu/~williams/mpfitexy/}.
Figure~\ref{fig_Delta_HR21}c,d show correlation plots between the low- and high- energy spectral indices. The results are somewhat different between the BPL and brmThick2 fits. Although for the BPL fit $\gamma_1$ and $\gamma_2$ show a significant correlation, the correlation between $\delta_1$ and $\delta_2$ is barely visible, implying that the low- and high- energy slopes in the spectrum of nonthermal electrons are independent. If so, the correlation between $\gamma_1$ and $\gamma_2$ can be understood as a result of some contribution of high-energy electrons into the low-energy photon spectrum.

The classical thick-target model \citep{Brown_1971, Somov_Syrovatskiy_1976} implies a simple unique relationship between the photon and electron spectral indices, $\delta=\gamma+1$. However, one can note that the numbers in Table~\ref{table_KW_Spectra} do not follow this simple formula because additional physical processes are included in \textit{brmThick2} model and also because of potential complexity of the spectrum, which is not captured by the simplified models.
The two bottom panels of Figure~\ref{fig_Delta_HR21} display the correlation plots between the low-energy ($\gamma_1$ and $\delta_1$) and high-energy ($\gamma_2$ and $\delta_2$) spectral indices, respectively. In both cases there is a significant correlation, which, however, noticeably deviates from the expectation based on the classical thick-target emission model. Interestingly, for the high-energy spectral indices the regression curve is shallower than $\delta_2=\gamma_2+1$ (dashed line) although most of the index pairs are quantitatively consistent with the expectation of $\delta_2=\gamma_2+1$. In contrast, for the low-energy indices the regression is steeper and does not cross the data points at all, which  can again be interpreted in terms of contribution from the high-energy electrons to the lower-energy X-ray emission.
%\newpage

The spectra are hard in the entire considered energy range. The low-energy part of the spectrum shows  $\gamma_1 \sim 2-3.3 $ and $\delta_1 \sim 2-4$. The steeper high-energy spectra are still hard with $\gamma_2 \sim 2.7-3.3 $ and $\delta_2 \sim 3.7-4.1$.
The two last spectra, 7 and 8, measured at the decay phase of the
flare, have two peculiarities: a dip at $\simeq$250~keV
and a systematic excess of emission above $\simeq$500 keV as
compared to the model. %(see on-line figure set, Figure~\ref{fig_KW_spectra}).---are we going to prepare the on-line figure set? Excess compared with what model? Any? Or a single power-law? }

%%%%%%%%%%%%% Here we use Earth-crossing time%%%%%%%%%%%%%%%%%%%%%%%%%

The fluence of the flare measured from 01:34:54.105 UT to 01:35:13.817 UT is $(1.214 \pm 0.020) \times 10^{-4}$~erg~cm$^{-2}$ and the 64~ms peak flux measured from 01:35:00.761 UT to 01:35:01.273 UT is $(1.95 \pm 0.05) \times 10^{-5}$~erg~cm$^{-2}$~s$^{-1}$ (both in the 20--1000~keV range). Assuming isotropic emission, the corresponding radiated X-ray energy  is $(3.36 \pm 0.06) \times 10^{23}$~erg and the peak HXR luminosity is $(5.40 \pm 0.13) \times 10^{22}$~erg~s$^{-1}$.

%% E= 4*Pi*(Wind_distance)^2*Fluence. I've got E=2.10 * 10^23 erg in previous result
%$(2.13 \pm 0.02) \times 10^{23}$~erg and the peak HXR luminosity is $(3.96 \pm 0.25) \times 10^{22}$~erg~s$^{-1}$. Why??? %All errors are given at the 68\% confidence level.

%\newpage
%\begin{landscape}

\begin{deluxetable*}{ccc|cccccc|ccccc}
\tablecolumns{12}
\tablewidth{0pc}
\tabletypesize{\footnotesize}
\tablecaption{Summary of the \kw\ spectral fits (20 keV -- 1 MeV)\label{table_KW_Spectra}}
\tablehead{\colhead{} & \colhead{} & \colhead{} & \multicolumn{6}{c}{\textbf{BPL}} \\
\colhead{No.} & \colhead{T$_{start}$\tablenotemark{a}} & \colhead{$\Delta$T} & \colhead{$\gamma_1$} & \colhead{$\gamma_2$}  & \colhead{$E_{br,ph}$} & \colhead{A\tablenotemark{b}} & \colhead{Flux\tablenotemark{c}} & \colhead{$\chi^2$/dof} \\
\colhead{} & \colhead{(s)} & \colhead{(s)} & \colhead{} & \colhead{} & \colhead{keV} & \colhead{} & \colhead{} & \colhead{} }
\startdata

1 & 0.000 & 5.632 & 2.53$_{-0.12}^{+0.08}$ & 2.82$_{-0.08}^{+0.10}$ & 91$_{-28}^{+30}$ & 0.079$_{-0.004}^{+0.008}$ & 4.48$_{-0.13}^{+0.13}$ & 0.67(36.7/55)  \\
2 & 5.632 & 1.024 & 2.30$_{-0.07}^{+0.06}$ & 2.85$_{-0.10}^{+0.11}$ & 106$_{-15}^{+15}$ & 0.268$_{-0.011}^{+0.014}$ & 13.3$_{-0.3}^{+0.3}$ & 0.86(47.4/55) \\
3 & 6.656 & 0.512 & 2.14$_{-0.14}^{+0.09}$ & 2.75$_{-0.10}^{+0.13}$ & 94$_{-19}^{+23}$ & 0.424$_{-0.028}^{+0.048}$ & 19.5$_{-0.5}^{+0.5}$ & 0.82(29.4/36) \\
4 & 7.168 & 1.024 & 2.37$_{-0.10}^{+0.06}$ & 2.94$_{-0.13}^{+0.12}$ & 112$_{-29}^{+17}$ & 0.232$_{-0.009}^{+0.017}$ & 11.9$_{-0.3}^{+0.3}$ & 1.1(60.5/55) \\
5 & 8.192 & 1.024 & 2.29$_{-0.13}^{+0.13}$ & 2.95$_{-0.06}^{+0.08}$ & 68$_{-7}^{+14}$ & 0.33$_{-0.03}^{+0.04}$ & 14.6$_{-0.4}^{+0.4}$ & 0.87(47.8/55) \\
6 & 9.216 & 0.512 & 2.17$_{-0.24}^{+0.19}$ & 2.86$_{-0.07}^{+0.10}$ & 58$_{-8}^{+17}$ & 0.48$_{-0.08}^{+0.12}$ & 18.8$_{-0.6}^{+0.6}$ & 0.71(24.8/35) \\
7 & 9.728 & 1.792 & 2.62$_{-0.15}^{+0.12}$ & 3.10$_{-0.06}^{+0.07}$ & 58$_{-7}^{+8}$ & 0.186$_{-0.018}^{+0.027}$ & 9.9$_{-0.3}^{+0.3}$ & 1.15(47.0/41) \\
8\tablenotemark{f} & 11.520 & 8.192 & 3.24$_{-0.07}^{+0.07}$ & 3.24$_{-0.07}^{+0.07}$ & \nodata & 0.0200$_{-0.0010}^{+0.0010}$ & 2.21$_{-0.20}^{+0.20}$ & 0.82(46.8/57) \\
\enddata
%
%\startdata
%%%%%%%%%%%%%%%%%%%%%%%%%%%%%%%%%%%%%%%%%%%%%%%%%%%%%%%%%%%%%%%%%%%%%%%%%%%%%%%%
%here \kw\ time is replaced by Earth-crossing time (time of flight =0.231 s)%%%%
%%%%%%%%%%%%%%%%%%%%%%%%%%%%%%%%%%%%%%%%%%%%%%%%%%%%%%%%%%%%%%%%%%%%%%%%%%%%%%%%
%\tablenotetext{a}{Since $t_0$(KW)=5694.105 s UT (01:34:54.105).}
%%
%\tablenotetext{b}{In units of photons~cm$^{-2}$~s$^{-1}$~keV$^{-1}$.}
%%
%\tablenotetext{c}{In units of erg~10$^{-6}~cm^{-2}~s^{-1}$ in energy range 20--1000~keV.}
%%
%\tablenotetext{d}{In units of electrons~10$^{35}$~s$^{-1}$.}
%%
%\tablenotetext{e}{Fitted with frozen $E_{br,e}$.}
%%
%\tablenotetext{f}{Fitted by a single power law.}
\end{deluxetable*}
\setcounter{table}{0}
\begin{deluxetable*}{ccc|ccccc|cccccc}
%\nonumber
%
\tablecolumns{12}
\tablewidth{0pc}
\tabletypesize{\footnotesize}
\tablecaption{(Continued)}
\tablehead{\colhead{} & \colhead{} & \colhead{} &  \multicolumn{5}{c}{\textbf{brmThick2}}\\
\colhead{No.} & \colhead{T$_{start}$\tablenotemark{a}} & \colhead{$\Delta$T}  & \colhead{$\delta_1$} & \colhead{$\delta_2$}  & \colhead{$E_{br, e}$} & \colhead{Electron } & \colhead{$\chi^2$/dof}\\
\colhead{} & \colhead{(s)} & \colhead{(s)}  & \colhead{} & \colhead{} & \colhead{keV} & \colhead{flux\tablenotemark{d}} & \colhead{}}
\startdata

1 & 0.000 & 5.632 & 3.08$_{-0.42}^{+0.23}$ & 3.75$_{-0.09}^{+0.14}$ & 208$_{-72}^{+151}$ & 1.0$_{-0.5}^{+0.6}$ & 0.66(36.5/55)\\
2 & 5.632 & 1.024 & 2.67$_{-0.42}^{+0.25}$ & 3.77$_{-0.11}^{+0.16}$ & 247$_{-62}^{+105}$ & 1.1$_{-0.6}^{+0.8}$ & 0.92(50.6/55)\\
3 & 6.656 & 0.512 & 2.47$_{-0.62}^{+0.32}$ & 3.84$_{-0.16}^{+0.31}$ & 293$_{-93}^{+182}$ & 1.0$_{-0.6}^{+0.9}$ & 0.84(30.2/36)\\
4 & 7.168 & 1.024 & 2.77$_{-0.16}^{+0.15}$ & 3.93$_{-0.10}^{+0.11}$ & 282\tablenotemark{e} & 1.3$_{-0.4}^{+0.5}$ & 1.1(61.0/56)\\
5 & 8.192 & 1.024 & 2.50$_{-0.26}^{+0.23}$ & 3.94$_{-0.07}^{+0.08}$ & 169\tablenotemark{e} & 1.3$_{-0.5}^{+0.7}$ & 0.88(49.2/56)\\
6 & 9.216 & 0.512 & 2.2$_{-0.4}^{+0.3}$ & 3.84$_{-0.07}^{+0.08}$ & 144\tablenotemark{e} & 1.1$_{-0.5}^{+0.9}$ & 0.72(26.0/36)\\
7 & 9.728 & 1.792 & 3.23$_{-0.22}^{+0.20}$ & 4.01$_{-0.07}^{+0.03}$ & 144\tablenotemark{e} & 4.3$_{-1.5}^{+2.0}$ & 1.12(47.1/42)\\
8\tablenotemark{f} & 11.520 & 8.192 &  4.07$_{-0.07}^{+0.07}$ & 4.07$_{-0.07}^{+0.07}$ & \nodata & 4.1$_{-0.6}^{+0.8}$ & 0.81(46.4/57)\\
\enddata
%%%%%%%%%%%%%%%%%%%%%%%%%%%%%%%%%%%%%%%%%%%%%%%%%%%%%%%%%%%%%%%%%%%%%%%%%%%%%%%%
%here \kw\ time is replaced by Earth-crossing time (time of flight =0.231 s)%%%%
%%%%%%%%%%%%%%%%%%%%%%%%%%%%%%%%%%%%%%%%%%%%%%%%%%%%%%%%%%%%%%%%%%%%%%%%%%%%%%%%
\tablenotetext{a}{Since $t_0$(KW)=5694.105 s UT (01:34:54.105).}
\tablenotetext{b}{In units of photons~cm$^{-2}$~s$^{-1}$~keV$^{-1}$.}
\tablenotetext{c}{In units of erg~10$^{-6}~cm^{-2}~s^{-1}$ in energy range 20--1000~keV.}
\tablenotetext{d}{In units of electrons~10$^{35}$~s$^{-1}$.}
\tablenotetext{e}{Fitted with frozen $E_{br,e}$.}
\tablenotetext{f}{Fitted by a single power law.}
\end{deluxetable*}
%\end{landscape}

\subsubsection{\Mw\ Spectra}

To obtain the \mw\ spectral evolution we employed two synthetic dynamic spectra, `slow' (Figure~\ref{fig_over_20020310}b) and `fast' (Figure~\ref{fig_over_20020310}c), described in Section~\ref{S_data_overview}. The slow dynamic spectrum allows for longer tracking the burst evolution, while the fast one has an advantage of having a better time resolution and 80~GHz data that generally help to better constrain the high-frequency spectral slope. It has to be kept in mind, however, that if the spectral peak is too high ($\sim 35$~GHz) and the high-frequency slope is only constrained by {the 4~s smoothing of} the poorly defined 80~GHz data point {(see Figure~\ref{fig_LC}c)}, the value of high-frequency \mw\ spectral index could not be determined reliably.

Both slow and fast dynamic spectra were sequentially fitted with a so-called \mw\ generic function proposed by \cite{Stahli1989SoPh}:
\begin{equation}
\label{Eq_mw_fit}
S=e^{A} f^{\alpha}\left[1-e^{-e^{B} f^{-\beta}}\right],
\end{equation}
where $f$ is the frequency in GHz, $A$, $B$, $\alpha$, and $\beta$ are the free fitting parameters, from which the relevant spectral parameters are computed. Specifically, the low-frequency spectral index $\alpha_{\rm lf} \equiv \alpha$, while the high-frequency spectral index  is $\alpha_{\rm hf} = \alpha-\beta$. The peak frequency and the flux density at peak frequency are calculated from the shape of function $S$ as described by  \cite{Nita_etal_2004}.
Following \cite{Nita_etal_2004} we used the corresponding built-in functionality of the \verb"ovsa_explorer" widget from the OVSA software available from the SSW distribution.

\begin{figure}\centering
%C:\MyProjects\ColdFlare10mar2002\fit_mw_20020310_r2.pro ::  c:\MyProjects\ColdFlare10mar2002\Spectra\Spectra_20020310_R2.ps; p.1
%
\includegraphics[width=0.94\columnwidth]{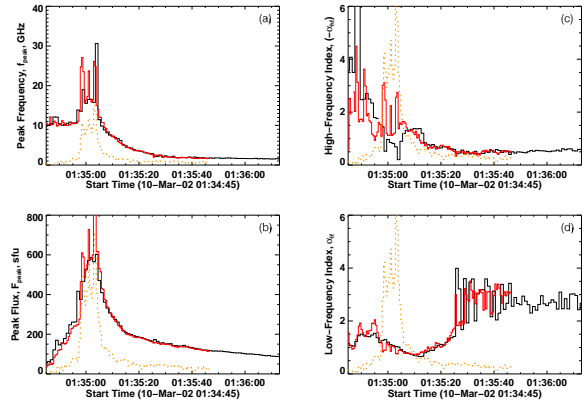}\\
\caption{\label{f_mw_spectral_fit_20020310} Microwave spectral fit parameter evolution: black curves---from 1~s data, red curves---from 0.5~s data, dotted orange curves are the appropriately scaled light curve at 35~GHz. (a)  peak frequency evolution. (b) peak flux evolution. (c) evolution of high-frequency spectral index.
 (d) evolution of low-frequency spectral index.
}
\end{figure}

Figure~\ref{f_mw_spectral_fit_20020310} shows a summary of the \mw\ spectral fit. The fit results obtained from the slow and fast dynamic spectra generally agree to each other, but nevertheless show a number of mismatches; especially, during the highly impulsive peak phase. The peak flux and peak frequency determined from the fast dynamic spectrum display a significantly stronger variation than the slow ones, which is real and reflects actual subsecond impulsiveness of the burst. On the other hand, the corresponding strong variations of the `fast' high-frequency spectral index are not real: they only take place when the spectral peak frequency is about 35~GHz, so that the high-frequency slope is only constrained by poorly known 80~GHz data{, as pointed out above}.

There are a number of  properties of the fit parameter evolution, which are noteworthy to mention. The spectral peak frequency demonstrates a remarkably large variation far more than by an order of magnitude: it is around 10~GHz at the burst rise phase, then it goes up to at least 35~GHz during the peak phase, and finally decreases down to roughly 1.5~GHz at the decay phase. Thus, the entire range of the spectral peak variation is within a factor larger than 20---a substantially broader range than for a `typical' microwave burst \citep{Nita_etal_2004, victor}. {This observation alone is a strong evidence that the magnetic field at the radio source at the decay phase is much smaller than that at the peak phase; see Section~\ref{S_Understand} for details.} Although there is an overall correspondence between the spectral peak flux and peak frequency in the sense that the larger the flux the larger the peak frequency, there is no perfect correlation between these two parameters. Indeed, if we compare the rise and decay phases, we note immediately that the same peak flux corresponds to substantially smaller peak frequency at the decay than at the rise phase. Note also, that at the early decay phase the peak flux and frequency decline highly consistently, while after roughly 01:35:30~UT the decrease of the peak frequency terminates although the peak flux continues to decline.

The high- and low- frequency spectral indices also display a substantial evolution. At the rise, peak\footnote{Neglecting the three outliers during three most impulsive peaks. }, and early decay phases the high-frequency spectral index shows a soft-hard-soft (SHS) pattern similar to that often reported for HXR spectra. But then, around 01:35:10~UT, the softening terminates and again gives a way to the spectral hardening. This hardening continues until 01:35:30~UT, when $\alpha_{\rm hf}$ hits the level of $\alpha_{\rm hf}\approx -1$ and than stays roughly constant. The low-frequency spectral index decreases all the way during the  rise, peak, and early decay phases until roughly 01:35:10~UT and then turns to increase until  01:35:30~UT, when it hits the level of $\alpha_{\rm lf}\approx 3$. After that it stays approximately constant at this level until the end of the event. Overall, the event demonstrates a strikingly prominent spectral variability over the rise, peak, and early decay phases (until $\sim$01:35:30~UT), while, in contrast,  shows no spectral evolution after that.

\subsubsection{Comparison of the X-ray and \Mw\ Spectral Indices}

\begin{figure}\centering
%C:\MyProjects\ColdFlare10mar2002\fit_mw_20020310_r2.pro ::  c:\MyProjects\ColdFlare10mar2002\Spectra\Spectra_20020310_R2.ps; p.2-3
%\qquad \qquad \qquad \qquad \qquad \qquad \qquad \qquad \qquad a\\
\includegraphics[width=0.43\columnwidth]{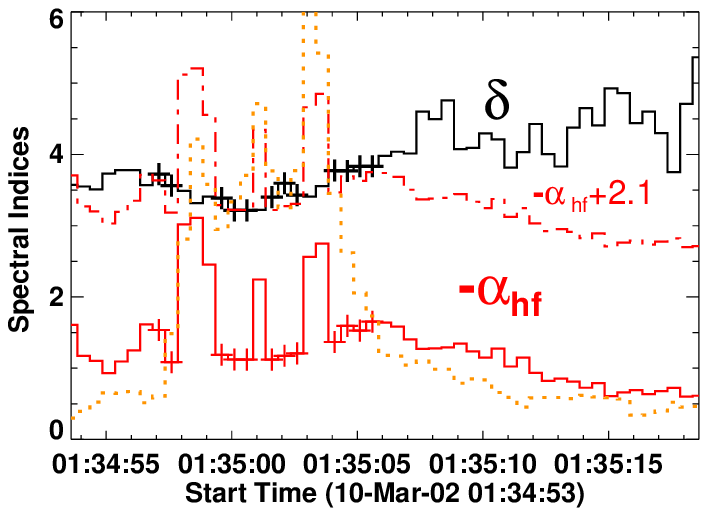}a
%\qquad \qquad \qquad \qquad \qquad \qquad \qquad \qquad \qquad b\\
\includegraphics[width=0.445\columnwidth]{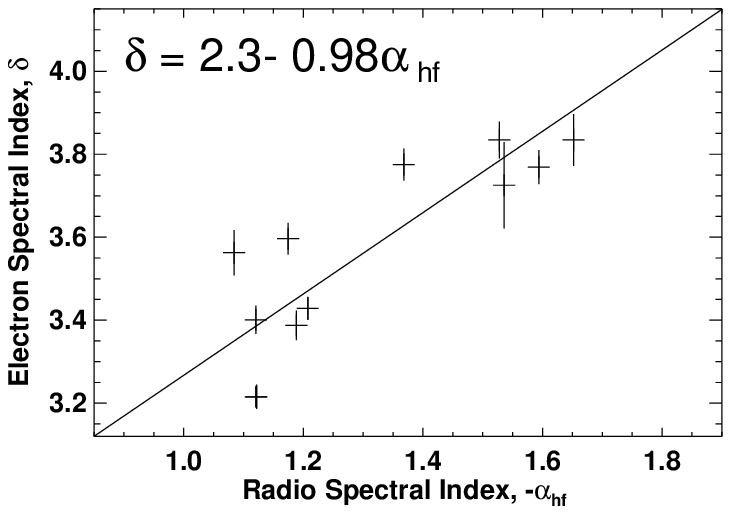}b\\
%\qquad \qquad \qquad \qquad \qquad \qquad \qquad \qquad \qquad c\\
%\includegraphics[width=0.43\columnwidth]{f_mw_X_ind_lag_corr_20020310.eps}\\
%\includegraphics[width=0.8\columnwidth]{goes_t.eps}\\
\caption{\label{f_mw_X_spectral_comp_20020310} Comparison between the \mw\ and HXR spectral indices. (a)  evolution of the electron flux spectral index $\delta$ (black curve) derived from the spectral hardness as explained in Section~\ref{s_kw_spec_fit} and \mw\ high-frequency spectral index $\alpha_{\rm hf}$ (red curve). The red dash-dotted curve shows the same \mw\ spectral index but displaced by 2.1 up to ease the visual comparison with the HXR-derived electron spectral index. The plus symbols show the data points  used for correlation analysis during the impulsive flare phase, which is envisioned by the 35~GHz light curve shown in orange dotted curve.   (b) the scatter plot of the indices and the corresponding linear fit. }
 %(c) the lag cross-correlation between the spectral indices. Apparently, the spectral evolution of the HXR and \mw\ emission show no time delay.

\end{figure}

Having both HXR and \mw\ spectral fits it is reasonable to compare the {`effective'} spectral index $\delta$ of the electron flux derived from the HXR {hardness ratio with Eq.~(\ref{Eq_hardness2delta})} and the high-frequency \mw\ spectral index $\alpha_{\rm hf}$ primarily defined by the energy spectrum of the fast electron number density in the source of the \mw\ emission. Figure~\ref{f_mw_X_spectral_comp_20020310}a shows that these spectral indices evolve consistently over the impulsive phase of the event as marked at the $\delta$ curve by plus signs outside the mentioned outliers. Both spectral indices show the SHS evolution, while they tend to disagree outside the impulsive peak. In spite of this apparent consistency, the scatter plot of the spectral indices during the impulsive phase displays no perfect correlation, Figure~\ref{f_mw_X_spectral_comp_20020310}b, although the data points roughly follow the linear regression law $\delta\approx 2.3-0.98\alpha_{\rm hf}$.

\subsection{Imaging}

\begin{figure}
%\epsscale{.99}
%KASHAPOVA
\includegraphics[width=0.47\columnwidth]{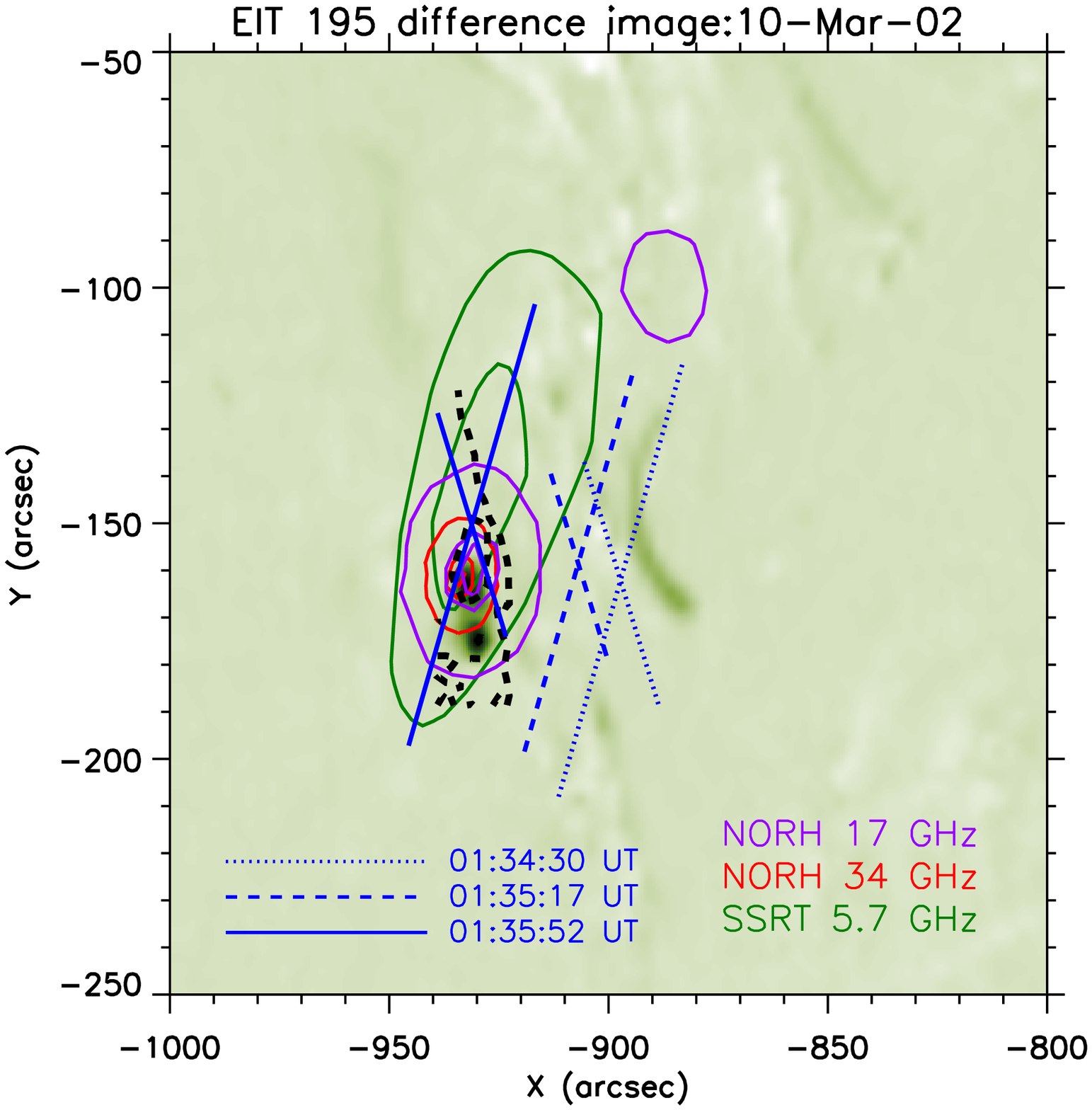}a
\includegraphics[width=0.47\columnwidth]{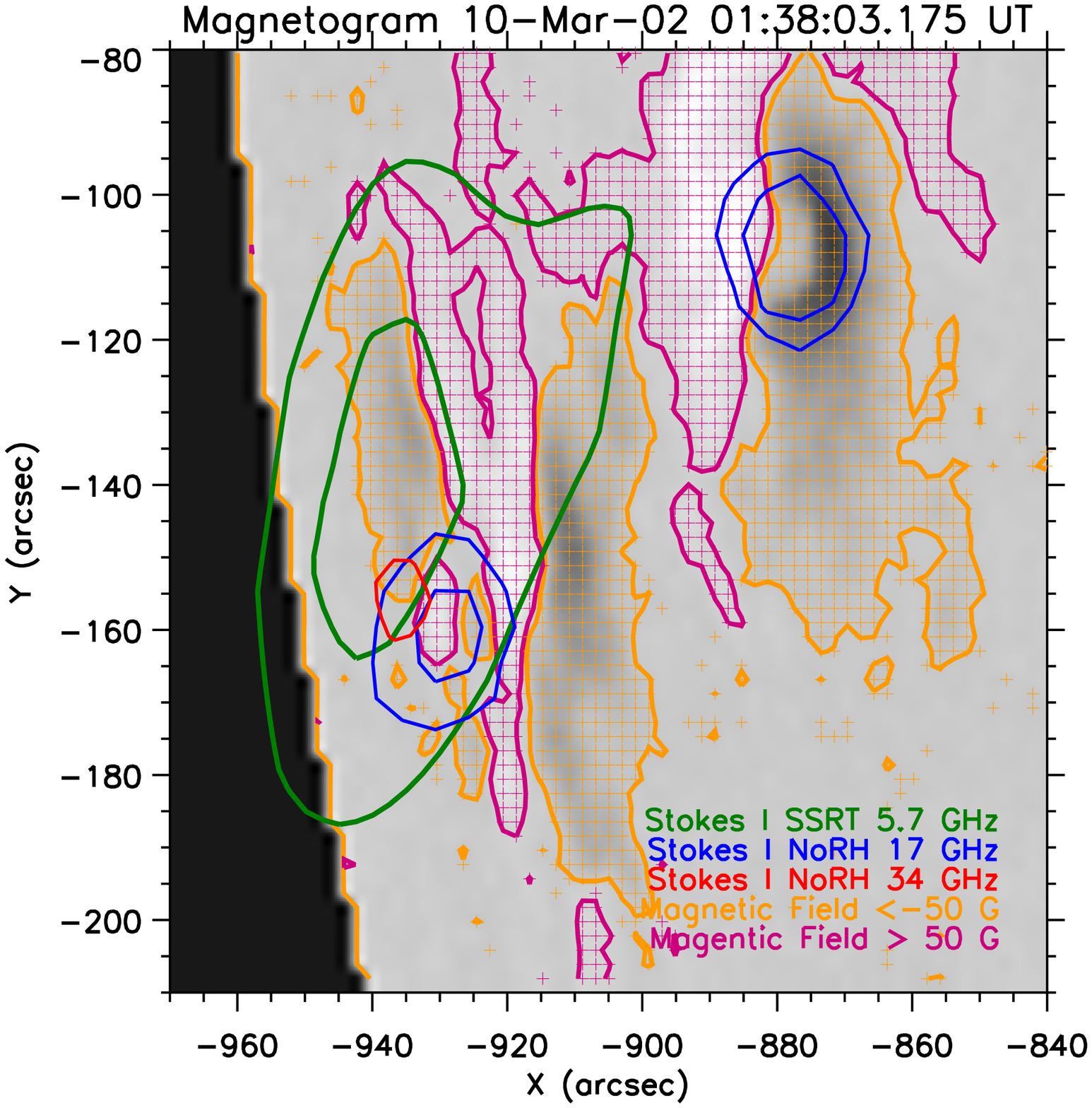}b %{ipa020310_013501V.eps} %{ipa020310_013803V.eps} %{ipa020310_013413V.eps} %{ipa020310_013632EIT.eps} %{ipa020310_013501EIT1.eps}
\caption{(a) The 195 \AA\ EIT/SOHO difference negative image between 01:36:14.7490~UT and 01:25:52.555~UT. %overlaid by contours of \mw\ emission.
Contours show \mw\ images at 34~GHz (red) and 17~GHz (violet) obtained at 01:35:01~UT by NoRH with intensity levels at 20\% and 70\% of the maximum. Green contours show 5.7~GHz image obtained by SSRT at 01:37:11~UT with intensity levels 70\% and 90\% of maximum. The dashed black line shows several neutral lines taken from photospheric magnetogram by MDI/SOHO. Intersections of the blue cross bars (dotted, dashed, and solid) present centroids of 5.7 GHz source at different moments. The direction of each cross bar shows the scanning direction of either EW or NS array. The length of each cross bar indicates the source width at half maximum over this bar direction. (b) SoHO/MDI magnetogram, where the locus of pixels with $B_{\|}>50$~G is shown in magenta, while with $B_{\|}<-50$~G is shown in orange. Contours of the \mw\ emission at 5.7, 17, \& 34~GHz are shown for the decay phase.
\label{figEIT}}
\end{figure}

A summary of the imaging data is given in Figure~\ref{figEIT}. The background color in Figure~\ref{figEIT}, left, shows one of the two 195~\AA~EIT/SOHO difference images\footnote{Three 195~\AA\ images were available taken on 01:25:52.555~UT, 01:36:14.749~UT, and 01:48:06.640~UT.} on which a representative set of microwave contours is superimposed along with the relevant neutral lines obtained from the SoHO/MDI photospheric magnetogram.

The difference image demonstrates the presence of two bright compact kernels at the flare time separated by about $10''$ in the North-South direction, which are not seen 12 minutes apart either before or after the flare. \Mw\ images at 17 and 34~GHz also show a double source structure but with a much larger separation between the sources, roughly $85''$. It is interesting that the EUV kernels are co-located with the strongest, southern, \mw\ source; the northern kernel spatially coincides with the centroid of the \mw\ source at the impulsive flare phase. There are two magnetic neutral lines; the shorter one separates the EUV kernels, while the longer one separates the \mw\ sources. These relationships imply that a magnetic connectivity is possible both between the EUV kernels and between the \mw\ sources.

\begin{figure}\centering
%ALTYNTSEV
\includegraphics[width=0.94\columnwidth]{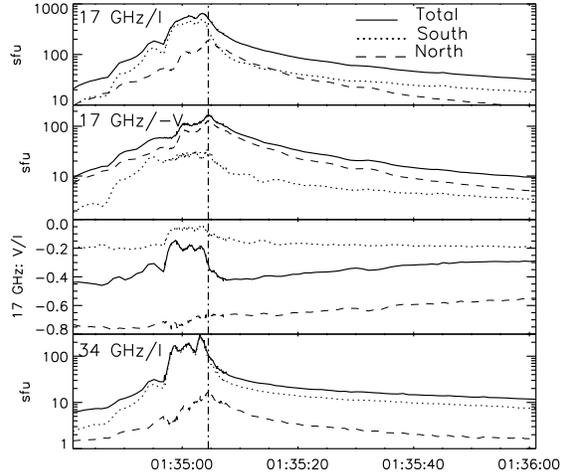}\\
\caption{\label{f_mw_sources_20020310} NoRH light curves: (a) flux densities from the north and south sources along with the integrated data at 17~GHz; (b) the same for the Stokes V parameters; (c) the same for degree of polarization; and (d) the flux densities at 34~GHz. %  Верх: Временные профили потоков излучения на 17 ГГц в из северного и  южного источников в сравнении с интегральным потоком вспышки.  Вторая панель: То же в потоках в круговой поляризации. Для удобства профили приведены с обратным знаком. Третья панель: Профили изменения степени поляризации.  Низ: Временные профили потоков излучения на 34 ГГц.
}
\end{figure}

The positions of the 17 and 34 GHz sources do not change at the course of the flare. At the impulsive phase the southern source dominates at both 17 and 34 GHz (Figure~\ref{figEIT}), while at the decay phase the brightness of these two sources is comparable at 17 GHz; however, the northern source is not seen at all at 34 GHz. Both \mw\ sources produce left circular polarization (LCP)  at 17~GHz. Remarkably (Figure~\ref{f_mw_sources_20020310}), the Northern source has a very strong degree of polarization, which reaches up to 80\% at the beginning of the rise phase and remains strong afterwards, $\sim 60\%$. In contrast, the degree of polarization of the stronger, Southern source is much weaker being about 20\% on average, but is strongly reduced during the impulsive phase, when the 34~GHz emission has the strongest intensity. {The spatially resolved \mw\ light curves display a prominent time delay between the southern and the northern sources, which implies that the electron acceleration takes place at or close to the southern source, while the electrons reach the remote northern source only after traveling  roughly 2~s through a coronal loop.}

The first 2D image at 5.7 GHz is available at 01:37:11 UT, i.e., already at the decay phase. There is one single source which is displaced compared with any of the high-frequency sources in a way implying a loop-like connectivity between the NoRH sources; its centroid is located closer to the southern than to the northern NoRH source. The 5.7 GHz source evolves noticeably during the flare. Given that SSRT produces only one 2D map per 2--3 minutes, we employ 1D SSRT scans to study this evolution. The intersections of the dotted, dashed, and solid blue lines in Figure~\ref{figEIT}a (from right to left) show the locations of the 5.7~GHz source centroid before, at, and after the impulsive phase of the event, respectively. Apparently, the source centroid moves eastward with the velocity estimated as $3.2\times10^7$~cm/s during the rise, peak, and early decay phases and then stops moving such as its position is almost precisely the location of the 2D image taken at 01:37:11~UT. The apparent source sizes remain roughly constant during the burst being about $50''\times100''$, which implies the true sizes to be about $40''\times90''$ or less.

\subsection{Summary of the Data}
\label{S_Understand}

There are two striking features in the flare under study. The first one is a prominent variation of the time profile impulsiveness---from highly impulsive HXR and high-frequency microwave light curves (particularly, at 35~GHz)---to more and more gradual microwave light curves at progressively lower frequencies and the SXR light curves. This makes it difficult to think of this event in terms of the conventional Neupert effect. Indeed, Figure~\ref{fig_LC}d,e show that the light curves at 35~GHz or, equivalently, the HXR ones, do not correlate with the SXR time derivative, but correlate tightly with the derivative of the microwave light curve at 3.75~GHz. This correlation could imply that the 3.75~GHz light curve represents the thermal response of the heated plasma. However, the SXR time derivative correlates well with the 3.75~GHz light curve itself, which would instead imply that the 3.75~GHz light curve is the most representative one for the nonthermal particle impact. Thus, the classification of the light curves onto thermal and nonthermal ones based solely on their impulsiveness becomes inconclusive in our event, which calls for a more detailed analysis of the observed relationships.

The second striking property, closely related to the first one, is a remarkable spectral evolution of the microwave burst. During the rise phase and the impulsive peak the \mw\ spectrum displays a conventional (inverted bell-shaped) form with the spectral peak frequency well correlated with the peak flux \citep[cf.,][]{Nita_etal_2004}. This correlation is indicative of the optical thickness effect in the spectral peak formation \citep{1985ARA&A..23..169D, victor}. Then, at a later, gradual phase the spectrum becomes essentially flat between 1 and 10 GHz. Overall, the spectral evolution can be characterized as a progressive low-frequency flattening over the course of the flare. This flattening can be understood if the \mw\ source is getting more and more nonuniform with time. However, this is only a part of the puzzle: over the gradual phase the degree of polarization is unexpectedly large at 3.75--5.7~GHz as well as at 17--35~GHz indicative of optically thin emission at these spectral ranges, while small at 1--2~GHz and 9.4~GHz as expected for the optically thick emission. This implies that two distinct nonuniform sources (presumably, loops) are involved.

\section{Modeling}

The first {of the two} loop{s implied by the data}, no doubts, produces nonthermal emission that dominates the impulsive component at high frequencies. The second loop produces a gradual emission, which can be either thermal or nonthermal, that dominates at the lower frequencies. \blank{Below we} {We carefully} addressed the question if this gradual emission is thermal or nonthermal \blank{in greater detail} {and rejected the thermal model; see Appendix~\ref{S_therm_model}}.

\subsection{Nonthermal model for the gradual flare component}

Given that we rejected the thermal model of the gradual emission component in the given flare we turn now to a nonthermal model. The most challenging for a nonthermal model is to explain, why the lower-frequency light curves have a longer decay constant than the higher-frequency ones; the property which holds all the way from 35~GHz down to 0.6~GHz. Indeed, the extended \mw\ emission is commonly ascribed to the fast electron fraction trapped in the coronal part of the flaring loop (often, the looptop) due to magnetic mirroring effect \citep[e.g.,][and many others]{Melnikov_1994, Meln_Magun_1998, Bastian_etal_1998, Lee_Gary_2000, 2000ApJ...531.1109L, Kundu_etal_2001, melnikov_etal_2002}. Then, the fast electrons lose their energy due to Coulomb collisions. They are also scattered into the loss-cone and are getting lost from the loop due to the Coulomb scattering. The characteristic time constants of the energy loss and angular scattering due to Coulomb collisions both increase with the electron energy; so the high-energy electrons survive longer in the coronal trap than the lower-energy electrons. The higher-energy electrons radiate at higher frequency; thus, the higher-frequency \mw\ emission is supposed to have a longer decay time than the lower-frequency emission. This behavior has been observed in many cases, but our event displays an exactly opposite trend.

To understand the likely causes of this unusual trend we refer to a seminal paper by \citet{1994SoPh..152..409L}, who thoroughly studied four X-class flares with unusually flat \mw\ spectra in the 1--20~GHz range; the X9.4 1991-Mar-22 flare from their sample demonstrates the closest resemblance to our event in terms of the light curve duration vs frequency. \citet{1994SoPh..152..409L} undertook a simplified 3D modeling with a dipole magnetic loop to interpret the observed properties of the \mw\ emission in their event sample. They found that for a dipole loop, which is big and nonuniform (has a reasonably high mirror ratio), the \mw\ spectrum can be remarkably flat in a rather broad spectral range, even from 1--20~GHz for sufficiently large magnetic loop. This conclusion is confirmed by a more sophisticated 3D modeling reported by \citet{Kuznetsov_etal_2011}.

In addition to the interpretation of the flat spectra at the flare peak phase, \citet{1994SoPh..152..409L} also offered an elegant and convincing scenario for the flare evolution, which naturally results in the observed `anomalous' behavior of the \mw\ light curves vs frequency.
Originally, at the impulsive rise phase, the \mw\ source is relatively compact and occupies a volume with a relatively strong magnetic field, which results in high spectral peak frequency, $f_{\rm peak} \gtrsim 20$~GHz. Then, the radio source expands to occupy a much bigger nonuniform loop\footnote{\citet{1994SoPh..152..409L} proposed an `inflated' magnetic loop, but a similar effect can be achieved if the fast electrons gradually fill bigger and bigger fraction of a large magnetic trap.}, and produces the flattest spectrum at the phase, when the source becomes the most nonuniform (i.e.,  magnetic field \blank{values cover} {varies over} the broadest range of values). Later, at the decay phase, the radio source `shrinks' toward the looptop, where the magnetic field is low, thus, resulting in a decrease of the spectral peak frequency toward 1~GHz at the decay phase. \citet{1994SoPh..152..409L} noted that the fast electron spectral softening (with the electron spectral index $\delta$ change from around 3 at the peak phase to 4 at the decay phase) can further improve the consistency of the model \mw\ spectral evolution with the observed one. It is highly likely that a very similar scenario happened in our event in spite of the fact that it is only a C5.1 GOES class flare, i.e., two orders of magnitude smaller than the X-class flares analyzed by \citet{1994SoPh..152..409L}. %; we return to this and other peculiarities of this event later.

\subsection{{The flare morphology} suggested by the data.}

Locations of and relationships between various EUV and \mw\ sources suggest that this flare belongs to Hanaoka morphological type \citep{1997SoPh..173..319H}, where the flare energy release is believed to be driven by interaction between a small compact loop, whose footpoints are highlighted by the EUV kernels, and a bigger loop, whose footpoints are highlighted by the high-frequency \mw\ sources at 17 and 34~GHz, while the coronal part of the loop is implied by the 5.7~GHz image. Remarkably, this two-loop configuration is supported by other available data. For example, there are neutral lines both between the EUV kernels and between the  \mw\ footpoints, which indicates that the corresponding magnetic connectivities are likely.

Independent evidence in favor of two distinct sources comes from the \mw\ polarization spectrum. Indeed the degree of polarization is very small at 1--2~GHz indicating  optically thick emission at these frequencies, but turns high at 3.75--5.7~GHz manifesting optically thin emission here. However, the degree of polarization is low again at 9.4~GHz, while once again high at 17--35~GHz. This behavior of the degree of polarization is entirely inconsistent with a single (even though spatially nonuniform) source \citep{Kuznetsov_etal_2011}, while requires two distinct radio sources with strongly differing magnetic fields.

It is reasonable to assume that a smaller loop has a larger magnetic field, while a larger loop has a weaker magnetic field and, therefore, can form an efficient magnetic trap. The \mw\ light curves support this idea. Indeed, the high-frequency light curve at 35~GHz, which is supposed to originate from a source with the strongest magnetic field, is highly impulsive (as well as all HXR light curves) and does not show any evidence of the fast electron trapping in a coronal loop. Thus, it is likely formed in the small loop with strong magnetic field and small mirror ratio, which makes the magnetic trapping inefficient. In contrast, the lower-frequency optically thin light curve at 3.75~GHz has a delayed tail such as if the fast electrons were injected from the small loop (or from an interaction region of these two loops) and then accumulated in the bigger loop. This casual relationship is further supported by close correlation between the time derivative of the gradual 3.75~GHz light curve and the impulsive 35~GHz and HXR light curves.

\subsection{Validating the model with 3D modeling}

Even though the outlined flare model seems plausible, the current state-of-the-art requires that it is quantitatively validated by a 3D modeling based on appropriate magnetic extrapolation as in \citet{Fl_etal_2011, Fl_etal_2013, Fl_Xu_etal_2016, Nita_etal_2015, Kuznetsov_Kontar_2015}. However, the modeling is substantially complicated in our case for the following reasons. Since two different loops with presumably different twists are involved in the flaring process, it is unlikely that they could be reproduced within either potential or linear force-free field (LFFF) extrapolation used in the cited studies addressing a single loop only, so a nonlinear force-free field (NLFFF) extrapolation is called for. However, there is no vector magnetogram available to perform NLFFF extrapolation. We do have the line-of-sight magnetic data from SoHO/MDI, which is formally sufficient to produce a LFFF extrapolation, but it will necessarily be imperfect since the region of interest is located very close to the limb\footnote{Typically, a model built from a LFFF extrapolation close to the limb requires that the model magnetic field is scaled by a small number.}. This implies that we can only perform a number of tests with the available data, but not a comprehensive modeling. In particular, we are forced to create two separate LFFF models, with presumably different force-free parameters $\alpha$---one for each of the two loops involved in our flare.

With this reservation, we are going to employ the powerful GX Simulator tool \citep{Nita_etal_2015} to test if there is a model (a set of two different data cubes---one for each loop) consistent with photospheric magnetogram that can answer the following key questions about the event:
\begin{enumerate}
  \item  If the implied small and big loops can be reproduced in LFFF extrapolated data cubes and what $\alpha$ are needed for that?
  \item  Is it possible to populate the small loop with a distribution of fast electrons, which is consistent with the HXR data and, at the same time, capable of reproducing the high-frequency \mw\ spectrum?
  \item  Is it possible to populate the big loop with a distribution of fast electrons consistent with the HXR data to reproduce the low-frequency \mw\ spectrum?
  \item Could the entire spectrum be reproduced by the two-loop model?
  \item  Is it possible to get the LCP polarization from both 17~GHz sources?
  \item  Is it possible to get  a very high degree of LCP polarization from the north 17~GHz source?
  \item Could the entire polarization spectrum be reproduced by the two-loop model?
\end{enumerate}

\begin{figure} %\centering
\includegraphics[width=0.46\columnwidth,clip]{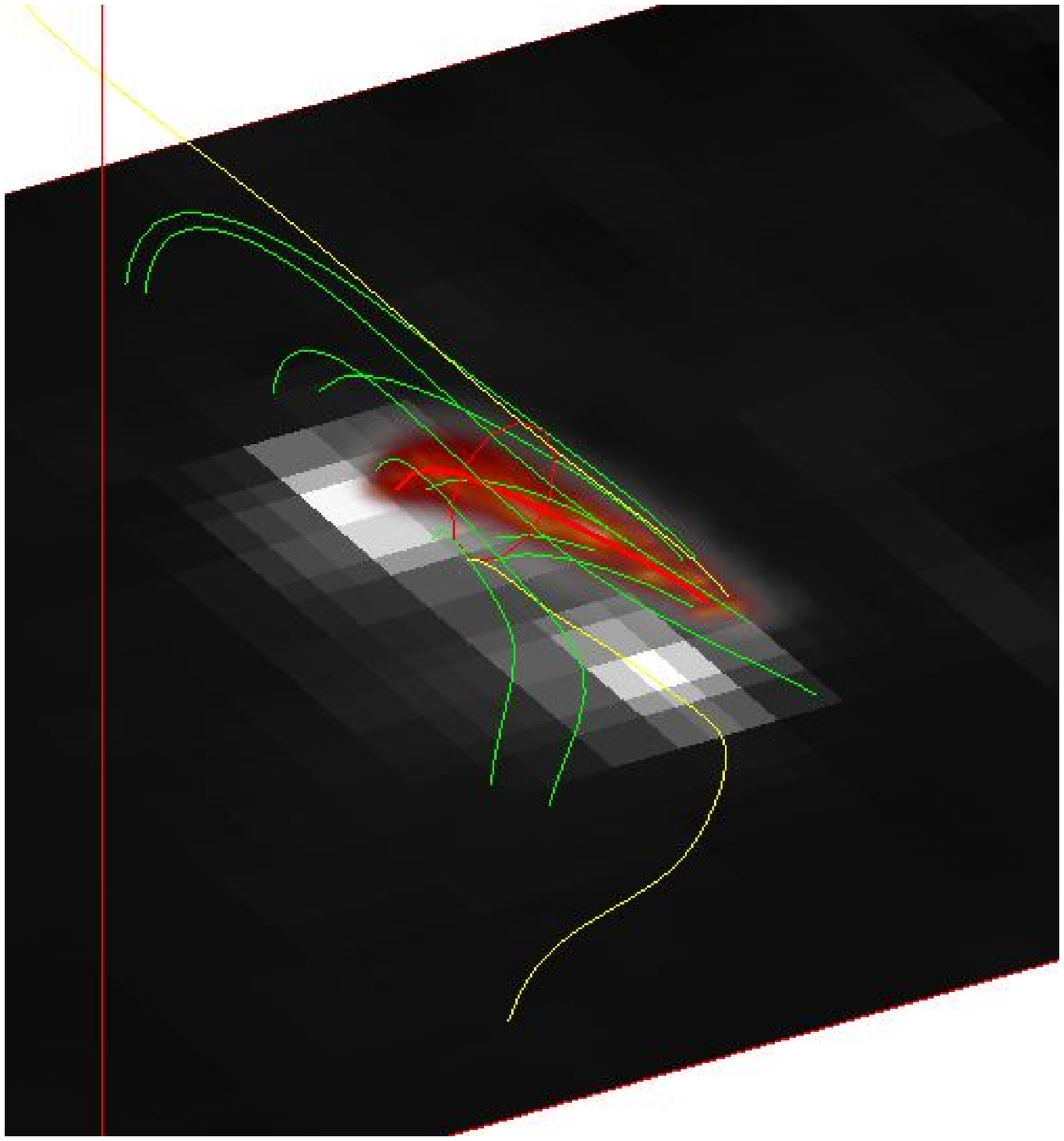}\
\includegraphics[width=0.5\columnwidth,clip]{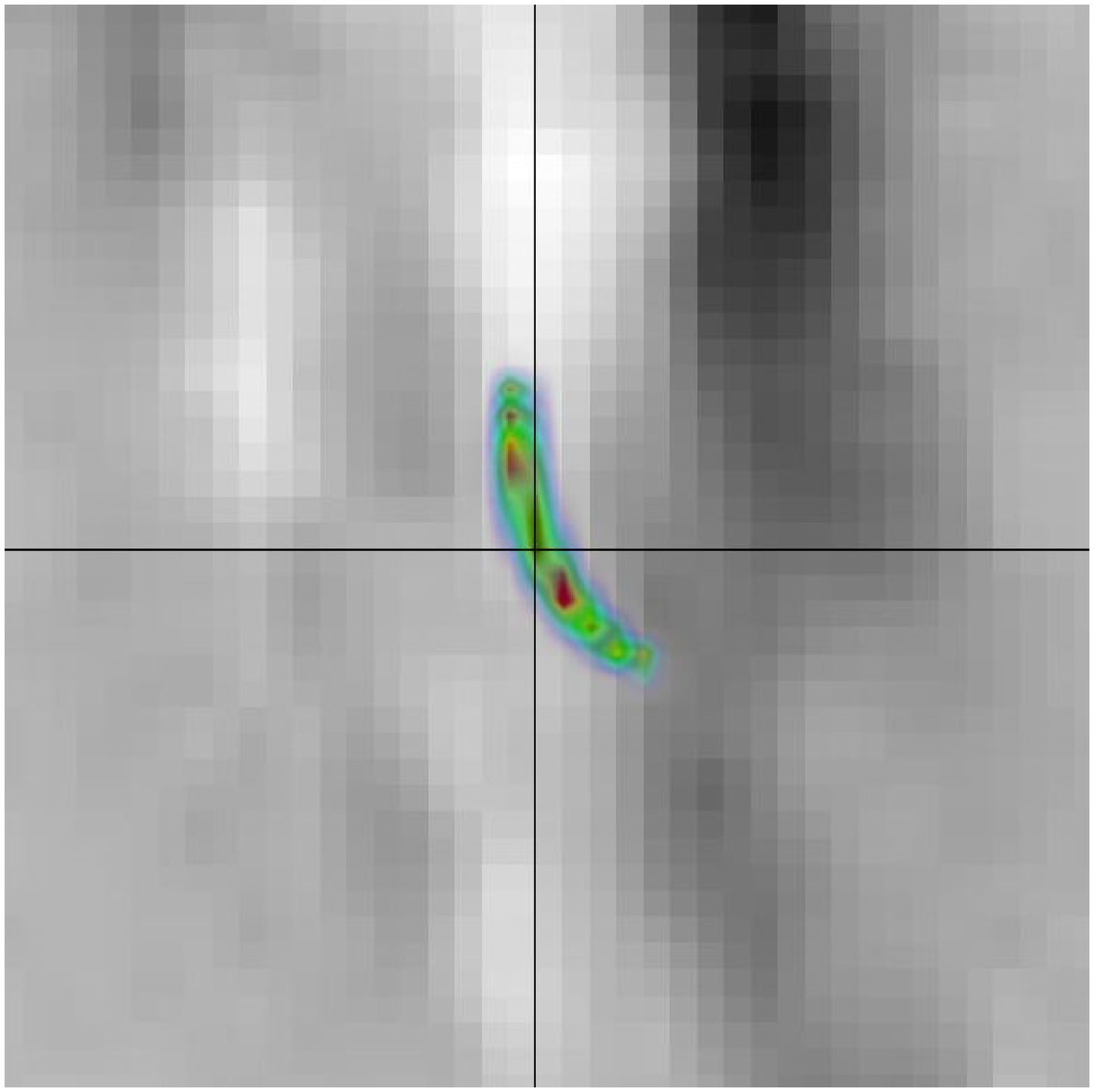}
\caption{A small loop connecting two 195 \AA\ kernels built out of the LFFF extrapolation with $\alpha\approx-1.75\times10^{-9}$~cm$^{-1}$. Left: perspective view with the EUV difference background image, magnetic structure visualized with a few closed (green) and open (yellow) field lines, cental field line of the flaring flux tube (red), and thermal electron density distribution (diffuse dark red volume). Right: number density of the fast electrons (diffuse green volume) on top of the LOS MDI magnetogram; top view.
\label{F_small_Loop_model}
}
\end{figure}

Let us start from a model needed to reproduce the small loop. After a number of trials and errors \citep[cf.][]{Fl_Xu_etal_2016} with the built-in LFFF engine of GX Simulator \citep{Nita_etal_2015}, we obtained a narrow range of $\alpha\approx-1.75\times10^{-9}$~cm$^{-1}$ with which the connectivity between the EUV kernels can be reproduced as shown in Figure~\ref{F_small_Loop_model}. The central field line (shown in red) has a length of $L_{\rm small}\approx 8.84\times10^8$~cm, the magnetic field value at the loop top\footnote{A scaling factor of 3 has been applied to the originally extrapolated magnetic data cube, \citep[cf.][]{Fl_Xu_etal_2016}.} $B_{\rm small,lt}\approx 620$~G, and the mirror ratio less than two.

\begin{figure} %\centering
%C:\MyProjects\ColdFlare10mar2002\Models/cf_20020310_spec_pol.pro
%\includegraphics[width=0.3\textwidth,clip]{f_small_loop_GX17_NoRH_195.eps} %\qquad
\includegraphics[width=0.3\columnwidth,clip]{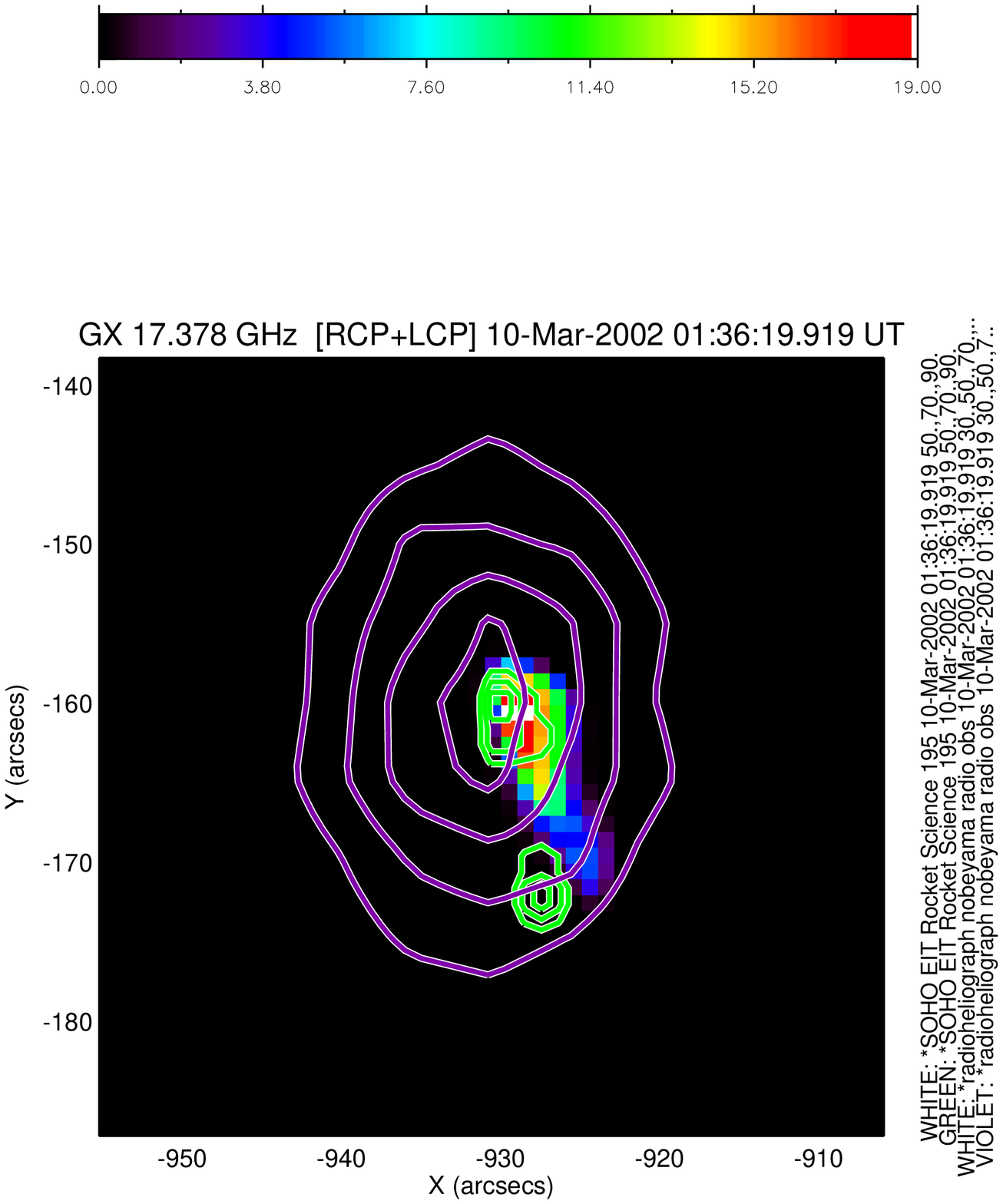}
\includegraphics[width=0.67\columnwidth,clip]{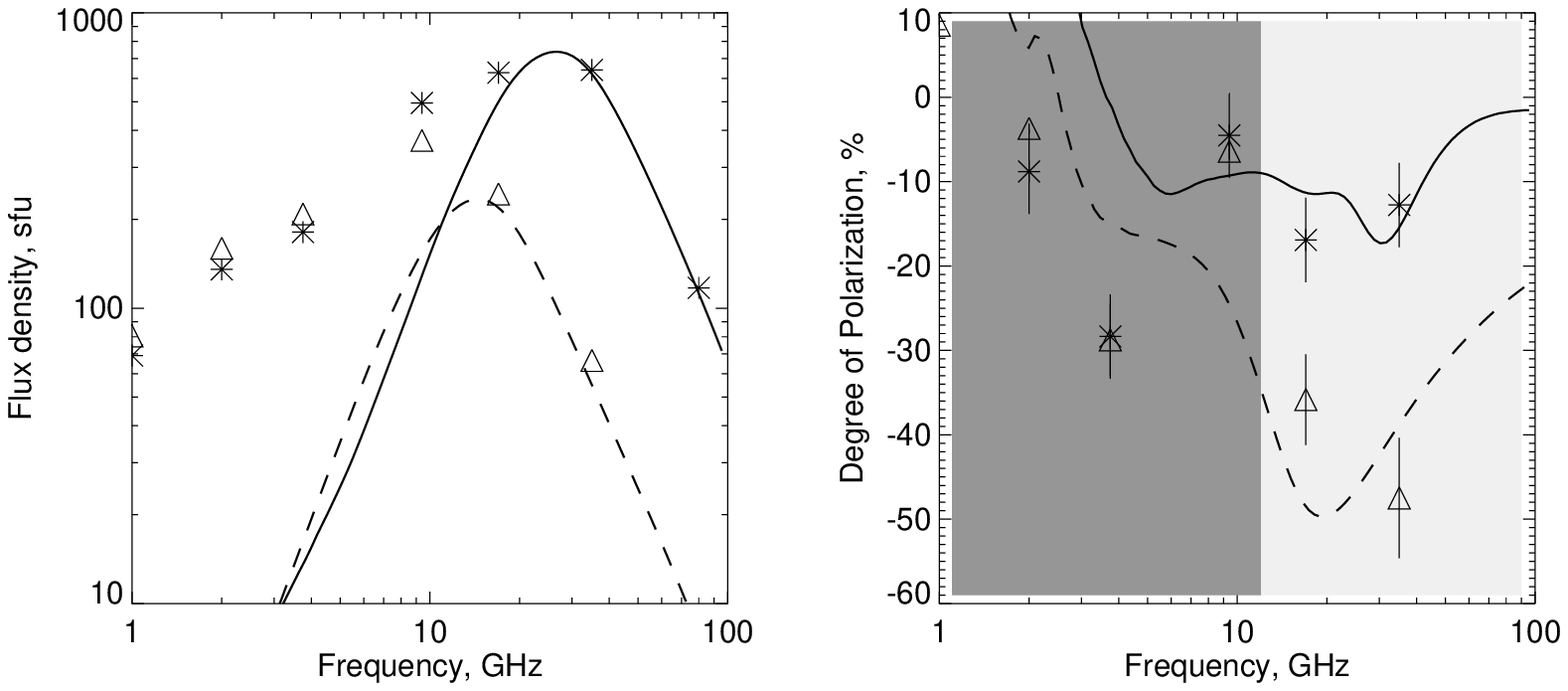}
\caption{Key outcome of the small loop modeling of high-frequency emission (above 10~GHz). Left: synthetic \mw\ image at 17~GHz computed from the adopted model along with EUV 195~\AA\ difference image (green contours) and NoRH 17~GHz image of the south source (violet contours). Middle: observed total power spectrum at the \mw\ peak time (01:35:03.600~UT, asterisks) and early decay phase (01:35:08~UT, triangles) and the corresponding model spectra from a beam-like (solid line) and isotropic (dashed line) angular distributions. Right: observed degree of polarization at the same times and the corresponding polarization spectra for a beam-like (solid line) and isotropic (dashed line) angular distributions. Only the area on top of the light grey background is relevant for the model-to-data comparison.
\label{F_small_Loop_model_MW}
}
\end{figure}

For \mw\ spectral modeling we select the peak time at  01:35:03.600~UT. One of the key ingredients for the modeling is the shape of the distribution function of the nonthermal electrons. Although the \mw\ data themselves can be successfully fit by a single power-law distribution of fast electrons over energy,\footnote{This is because the \mw\ spectrum at these high frequencies is not sensitive to the exact shape of the nonthermal electron spectrum at low energies, where the break of this spectrum is suggested by the \kw\ data. This results in a well-known uncertainty while estimating energy contents and other measures determined by the low-energy part of the nonthermal electron spectrum.} we adopt here a broken power-law as suggested by the \kw\ X-ray fit (cf. Table~\ref{table_KW_Spectra}).   We get a reasonably good \textit{spectral} match at high \mw\ frequencies if we populate this magnetic loop with fast electron distribution (cf. Table~\ref{table_KW_Spectra}, intervals 5-6)  with $E_{\min}=10$~keV, $E_{\rm br}=150$~keV, $E_{\max}=1.8$~MeV, $\delta_{r,1}=2.5$, $\delta_{r,2}=3.5$, and $N_{\rm r,tot}\sim10^{34}$ electrons distributed roughly uniformly over the loop length and isotropically over the pitch-angle. The thermal number density has almost no effect on this high-frequency emission: we varied the number density from $10^{10}$ to $5\times10^{11}$~cm$^{-3}$ with essentially no modification of the spectrum. The \textit{sense} of polarization corresponds to LCP wave at the high frequencies in agreement with observations, but the \textit{degree} of polarization is much stronger than observed. The degree of polarization can be reduced by either a tangled magnetic field structure at the source or having a beam-like anisotropy of the accelerated electrons \citep{Fl_Meln_2003b}. We investigated possible effect of the beam-like anisotropy in our case and found that it offers a much better match (solid curves in Figure~\ref{F_small_Loop_model_MW}) to the measured degree of polarization at  the impulsive peak than the isotropic distribution (dashed curves in Figure~\ref{F_small_Loop_model_MW}). The best fit is obtained for the  number density of the fast electrons $n_{\rm r}=5.2\times10^8$~cm$^{-3}$ (this is the peak value of the spatially nonuniform electron distribution) that corresponds to the total number of fast electrons at the source $N_{\rm r,tot}\approx1.35\times10^{34}$. %ne 7e8
Note, that the electron acceleration rate determined from the X-ray fit is about $1.2\times10^{35}$~electron/s, which implies that the electron escape time $\tau_{\rm esc}$ from the loop is roughly 0.1~s, which is three times larger than the time of flight ($L_{\rm small}/c\sim30$~ms) estimated for our loop length $L_{\rm small}\sim9\times10^8$~cm. Given the electron distribution is found to be beamed along the field lines, while the mirror ratio in this loop is small, $\sim2$, a more reasonable estimate for the escape time would be within 30~ms; this upper limit for the escape time is also implied by absence of any measurable (within 0.1~s time resolution) delay between the 35~GHz light curve and HXR light curves. Our two-loop model offers a natural solution for this discrepancy: with the numbers above we conclude that in fact the acceleration rate is roughly two times larger than that derived from the HXR fit, but the remaining ($\sim50\%$ of) accelerated electrons escape to the second, big loop,\footnote{A fraction of the nonthermal electrons can also escape along the open field lines.} rather than precipitate to the small loop footpoints; thus, they do not contribute to the HXR emission.

Apparently, we can estimate the escape time from the 35~GHz light curve decay time $\tau$, which is roughly 1~s, in even stronger contradiction with 0.03~s determined above. A reasonable way to reconcile this contradiction is to ascribe the decay segment of the light curve to the residual electron component trapped in the small loop. If so, this  residual component must be substantially more isotropic than the original beamed distribution at the impulsive acceleration phase. The implied evolution of the pitch-angle distribution from a beamed to a more isotropic or loss-cone one must leave a fingerprint in the corresponding evolution of the degree of polarization. Indeed, as we noticed in Section~\ref{S_light_curves}, see Figure~\ref{fig_LC}f, the degree of polarization goes up at 17 and 35~GHz at the early decay phase. It is interesting that the polarization data at this decay phase are quantitatively consistent with the isotropic distribution of radiating electrons. Indeed, the  model degree of polarization (dashed curve)  agrees well with the  data shown with the triangles  in Figure~\ref{F_small_Loop_model_MW}, right. The model assumes  the number density of the fast electrons $n_{\rm r}=6\times10^7$~cm$^{-3}$ that corresponds to the total number of fast electrons at the source $N_{\rm r,tot}\approx8\times10^{32}$, and  the same spectral parameters as at the peak phase. The corresponding observed (triangles) and model (dashed curve) flux densities are shown in the middle panel of Figure~\ref{F_small_Loop_model_MW}.

\begin{figure} %\centering
\includegraphics[width=0.55\columnwidth,clip]{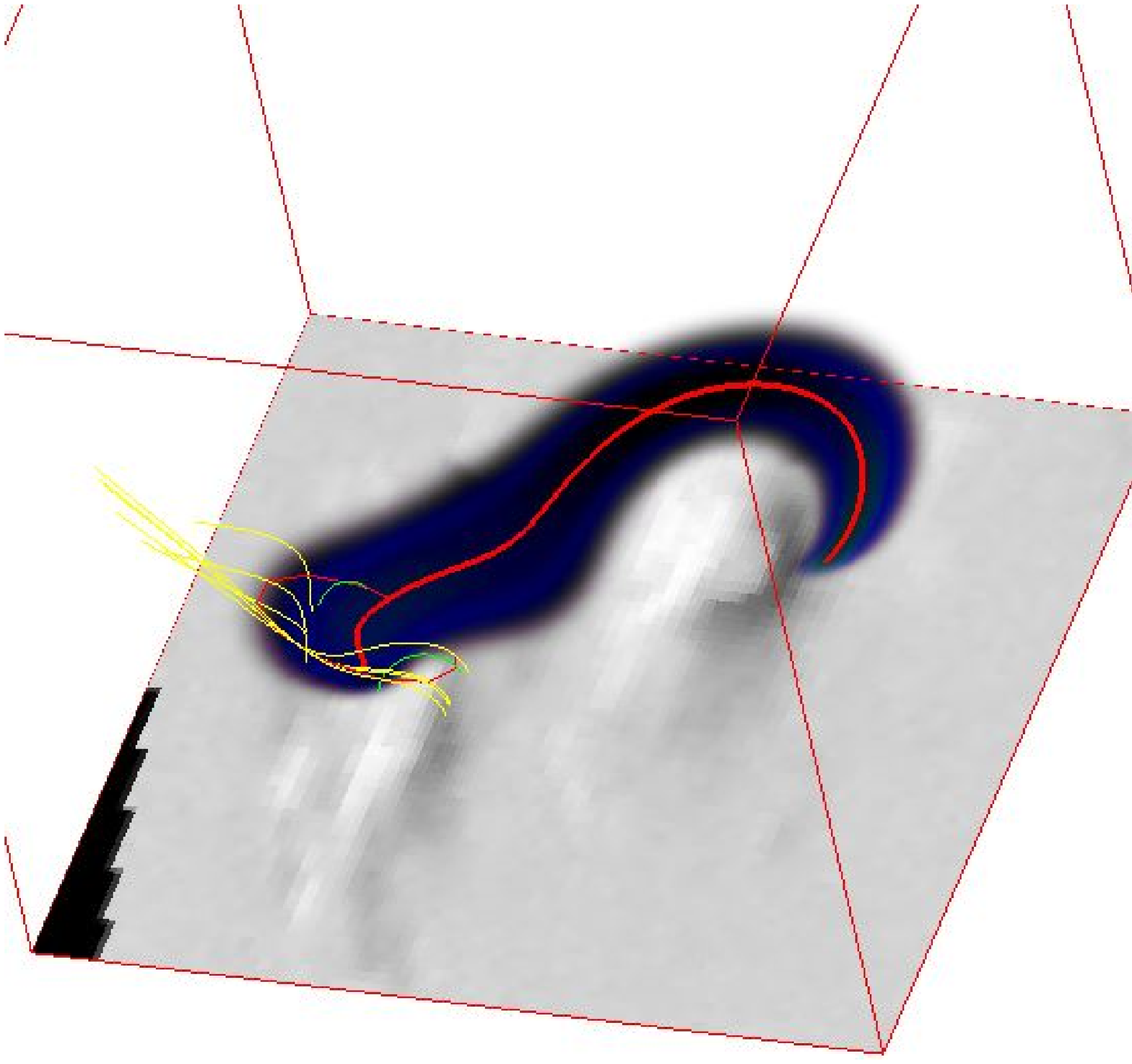}\qquad
\includegraphics[width=0.37\columnwidth,clip]{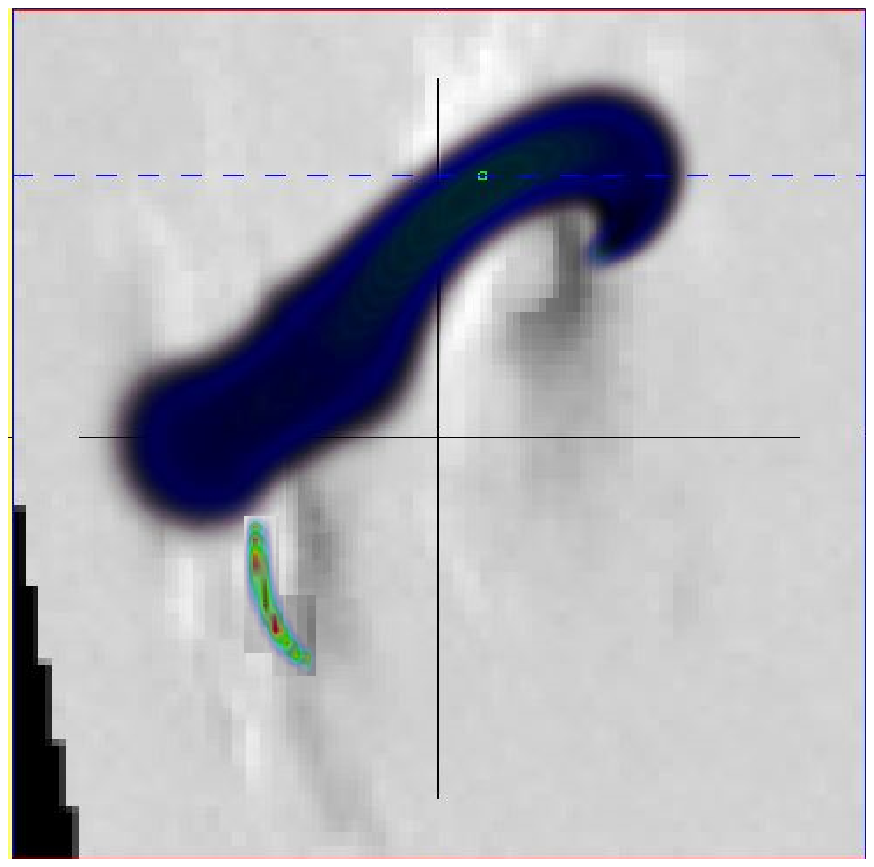}
\caption{A big loop connecting two NoRH 17~GHz sources (northern and southern) built out of the LFFF extrapolation with $\alpha\approx1.16\times10^{-9}$~cm$^{-1}$. Left: perspective view with the MDI LOS magnetogram background image,  cental field line of the flaring flux tube is shown in red, and the nonthermal electron density distribution shown in diffuse dark  volume. Right: top view with the small loop (green volume) added in scale.
\label{F_big_Loop_model}
}
\end{figure}

Let us turn \blank{now} to the big loop modeling. We use now a bigger field of view covering both northern and southern \mw\ sources. Building a big loop in the right place is possible for extrapolated data cubes with positive $\alpha$ centered around the value $\alpha\approx1.16\times10^{-9}$~cm$^{-1}$. The center field line of the model loop has the length  $L_{\rm big}\approx 8.2\times10^9$~cm, the magnetic field value at the loop top\footnote{A scaling factor of 0.58 has been applied to the originally extrapolated magnetic data cube.} $B_{\rm big,lt}\approx 30$~G, and the mirror ratio about four, see Figure~\ref{F_big_Loop_model}. We note that the nonthermal electron distribution in the big loop is poorer constrained than that in the small loop, because the accelerated electron distribution, before arriving at the big loop, is  modified by the energy-dependent escape time from the small loop/acceleration region and then by the energy-dependent trapping time in the big loop. Moreover, we have only a limited information about the high-frequency slope of the \mw\ emission from the big loop from the spatially resolved NoRH data, while the total power NoRP data are dominated by the small loop as explained above. Given all these uncertainties, for the big loop we adopt a single power-law energy spectrum of the nonthermal electrons with the same (high-energy) spectral index as for the small loop, $\delta_{r}=3.5$.

\begin{figure} %\centering
%C:\MyProjects\ColdFlare10mar2002\Models/cf_20020310_spec_pol.pro
\includegraphics[width=0.28\columnwidth,clip]{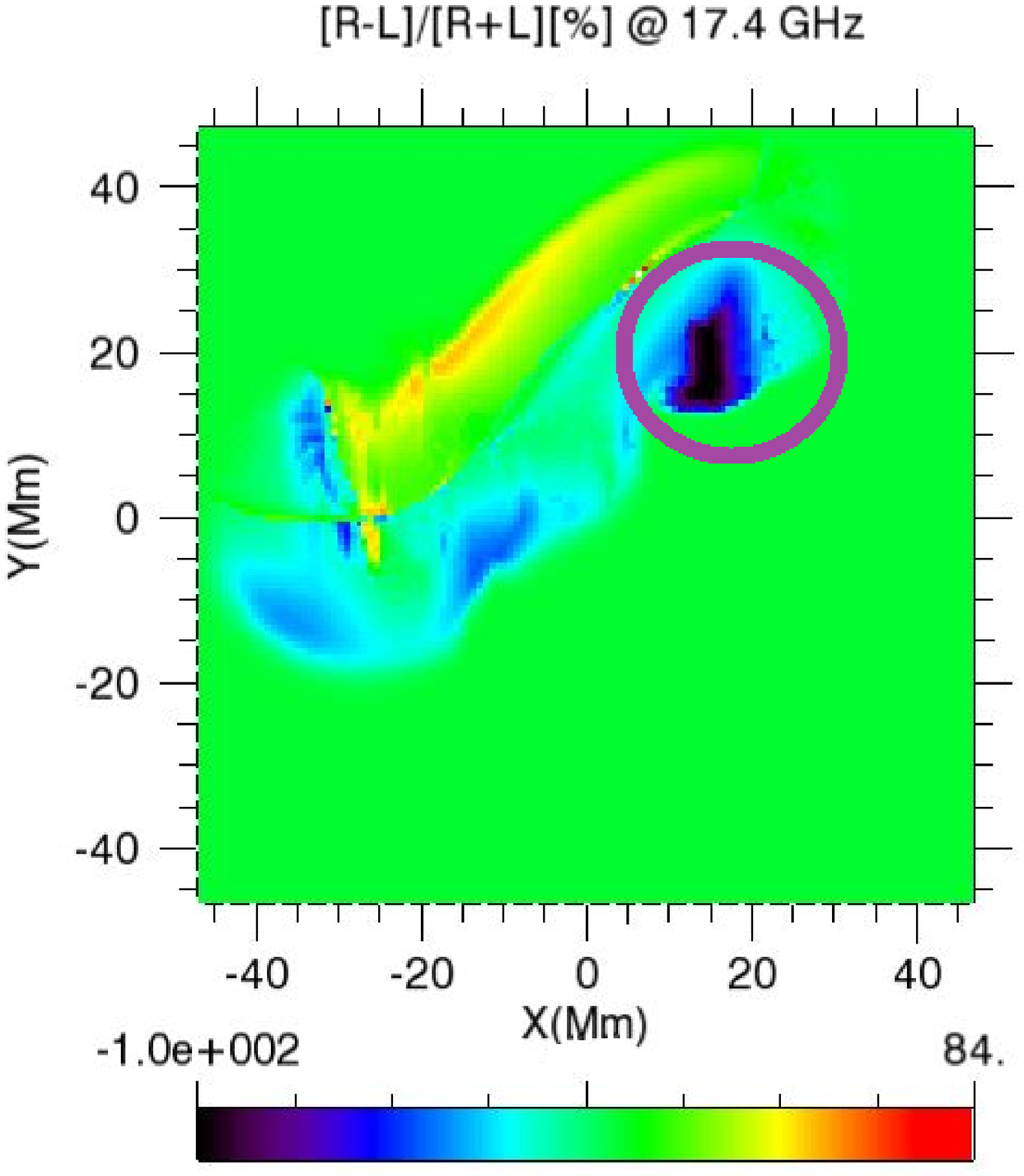} %\qquad
\includegraphics[width=0.69\columnwidth,clip]{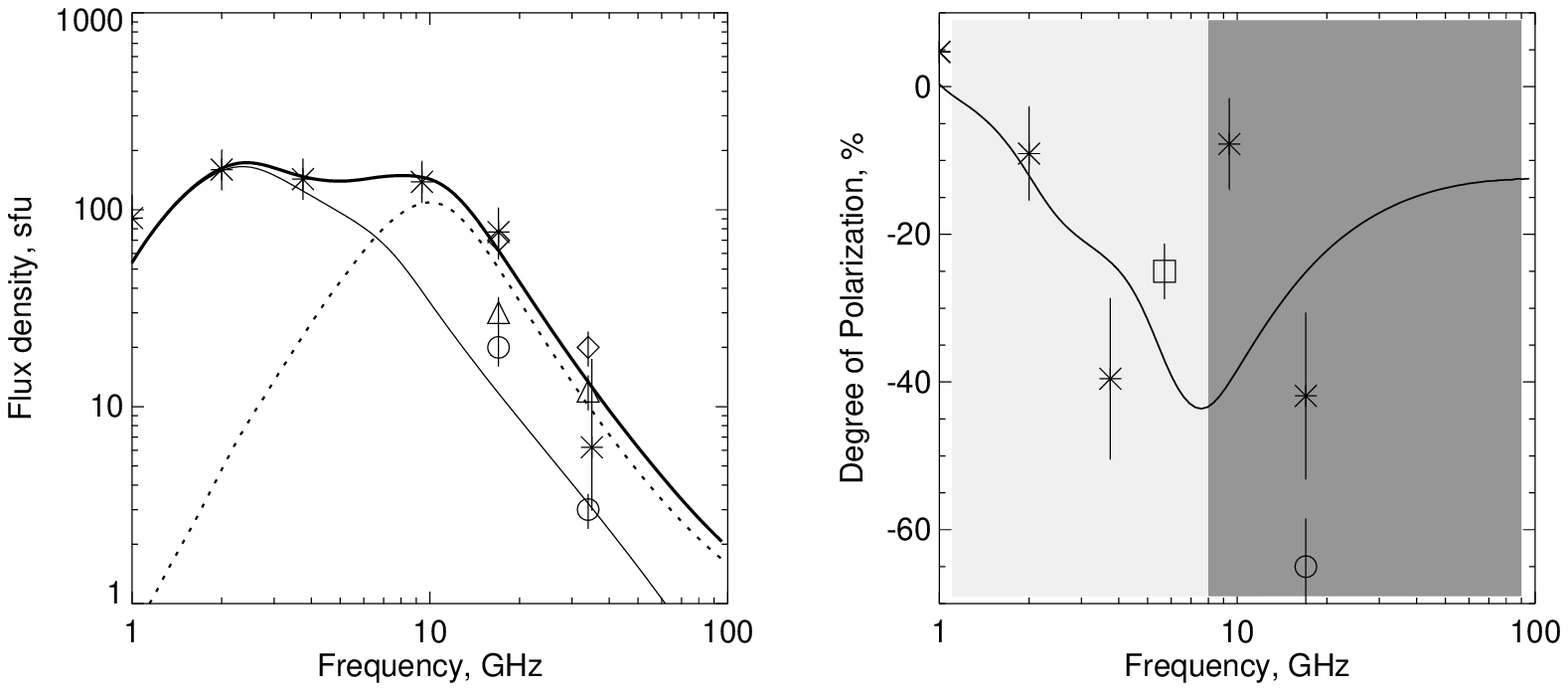}
\caption{Key outcome of the big loop modeling. Left: synthetic \mw\ polarization image at 17~GHz computed from the adopted model along with NoRH 17~GHz highly polarized Northern source (violet contour). Middle: observed total power spectrum at the decay phase (01:35:24.500~UT; asterisks) and spatially resolved NoRH data at 17 and 34~GHz from the northern source (open circles), the southern source (triangles), and total (diamonds) along with the corresponding model spectra from  the big loop (thin solid line),  the small loop  (dotted line), and sum of them (thick solid line). Right: observed degree of polarization at the decay phase (01:35:24.500~UT) from NoRP (triangles), SSRT at 5.7~GHz (square), and northern NoRH source at 17~GHz (open circle)  and the corresponding synthetic polarization spectrum produced by the big loop (solid line). Only the area on top of the light grey background is relevant for the model-to-data comparison.
\label{F_big_Loop_model_MW}
}
\end{figure}

For spectral modeling of the big loop contribution we select the time  01:35:24.500~UT at the decay phase---rather close to the end of the prominent spectral evolution, where emission from the big loop presumably dominates the \mw\ spectrum. We get a reasonably good spectral match at low frequencies if we populate this magnetic loop with fast electron distribution within the energy range starting from the same $E_{\min}=10$~keV in agreement with both HXR data and the small loop model to $E_{\max}=5$~MeV,\footnote{Although the value of $E_{\max}$ is poorly constrained by data, smaller $E_{\max}$ would result in a progressively stronger underestimation of the flux density at 34~GHz from the northern source shown by open circle in Figure~\ref{F_big_Loop_model_MW}.}  and the number density $n_{\rm r}=1.6\times10^7$~cm$^{-3}$ totaling in $N_{\rm r,tot}\approx5.7\times10^{34}$  electrons slightly concentrating towards the looptop, as expected due to particle trapping effect in the magnetic loops \citep[][]{melnikov_etal_2002}. The angular distribution is expected to have a loss-cone shape with the loss-cone angle $\theta_{\rm lt}=30^\circ$ in the top of the loop in agreement with the mirror ratio of four. In fact, the isotropic distribution was found to give the same results, so we give here the numbers relevant to the isotropic model, which is computationally faster than the anisotropic one. The thermal plasma density at the central field line of the big loop is adopted to be $n_0=5\times 10^9$~cm$^{-3}$. This model offers a very good match for the low-frequency part of the total power spectrum and also reproduces the correct level of the spatially resolved data from the northern NoRH source at 17~GHz and 34~GHz, although the flux density of the northern source at 17~GHz is slightly underestimated. Similarly, the model slightly underestimates the flux density at 1~GHz, which indicates that the real source has a slightly stronger nonuniformity than the model one \citep[cf. 3D models in][]{Kuznetsov_etal_2011}.

The model reproduces trends of the degree of polarization. Indeed, as it is seen from the right panel of Figure~\ref{F_big_Loop_model_MW}, the degree of polarization is close to zero at 1~GHz, while negative and relatively strong at 2~GHz and 3.75~GHz---all in agreement with observations. However,  values of the model degree of polarization at 1~GHz and 3.75~GHz deviate from the observed values by a factor of two. We varied the angular distribution of the fast electrons in the big loop, but this did not remove the mismatch of the polarizations. We conclude that the mismatch is likely due to imperfect geometry of the model loop compared with the real source (recall, the degree of polarization is highly sensitive to the viewing angle), which is very possible given the limitations of our modeling due to lack of constraints discussed in the top of this section.

Finally, we checked and confirmed that the combination of these two loops is capable of matching the entire \mw\ spectrum at a given time; to illustrate that we selected the same time frame 01:35:24.500~UT, see the thick solid line in Figure~\ref{F_big_Loop_model_MW}. Although the solution is far not unique, we can get a very good match to the entire spectrum between 1 and 35~GHz by adding emission (dotted line in Figure~\ref{F_big_Loop_model_MW}) from a small loop with the same central field line as in the impulsive phase, but having a somewhat broader spatial distribution of the accelerated electrons  transversely to this central field line.
A consistent fast electron distribution has the number density of the fast electrons $n_{\rm r}=1.5\times10^6$~cm$^{-3}$ that corresponds to the total number of fast electrons at the small loop $N_{\rm r,tot}\approx1.1\times10^{32}$  distributed isotropically with a modest concentration at the looptop \citep[cf.][]{melnikov_etal_2002}, and with the same broken power-law spectral shape. Given that we did not get the quantitative match of the degree of polarization at low frequencies, we did not try to match the entire polarization spectrum with the two-loop model.

With the described modeling we got a definitive `yes' for  questions 1--6 raised in the beginning of this section. The answer to the seventh question is quantitatively less definitive: we failed to quantitatively reproduce the degree of polarization at many frequencies or time frames analyzed. Nevertheless, we did obtain a generally right trend in all cases. Indeed, the sign of the degree of polarization is reproduced correctly for all time frames and at all frequencies; the model numbers are consistent with the observed ones within a factor of two. When we obtained a larger mismatch, we then found a much better match by adjusting the pitch-angle distribution of the fast electrons, which gave us additional important constraints on the model. We conclude that the proposed flare model based on two interacting loops is fully validated by the modeling performed here.

\section{Discussion}

We have described a puzzling cold flare event observed in X-ray domain by GOES and \kw\ and in \mw\ domain by a number of instruments in a rather broad spectral range, covering more than two orders of magnitude in frequency---from 0.6~GHz to 80~GHz. This flare displays a number of truly exceptional properties in all these domains. The HXR emission extends well above 1~MeV and it displays one of the hardest energy spectrum ever detected \citep{Vestrand1999}  for a C-class flares. In SXR the GOES light curves are substantially delayed compared to what is expected according to the Neupert effect. In \mw s there is a number of peculiarities. The spectral peak frequency displays an exceptional variation over the burst between at least 35~GHz at the impulsive peak down to 1.5~GHz at the decay phase---the property observed for some of the strongest events, but clearly at odds for a modest C-class flare. The \mw\ light curves show a remarkable diversity of their shapes---from very impulsive and highly correlated with HXR light curves at 35~GHz to more and more gradual at lower and lower frequencies---again, observed for a few X-class flares, but not for modest events. However, this exceptional spectral evolution takes place only over the impulsive phase and early decay phase, but then the spectral evolution switches off at some point and the microwave emission slowly decays with a constant spectral shape. In addition, during this remarkable spectral evolution stage, the coronal part of the \mw\ source moves rapidly eastward with the velocity of 320~km/s, but then stops and stays at the same location during the remaining decay phase.

The data analysis and 3D modeling  suggest that all remarkable properties of this event can quantitatively be understood within a model involving energy release due to interaction of two non-potential magnetic flux tubes---one small and one big with different twists ($\alpha\approx -1.75\times10^{-9}$~cm$^{-1}$ and $\alpha\approx 1.16\times10^{-9}$~cm$^{-1}$, respectively). Electrons are accelerated due to interaction (magnetic reconnection) between these two loops and then divided in roughly equal numbers between these two loops. The electrons injected into the small loop have a beam-like distribution directed towards the southern EUV kernel. {This finding about the beamed pitch-angle distribution of the nonthermal electrons is highly important for understanding the electron acceleration and transport, yet no routine diagnostics of the electron angular distribution is available. \citet{2006ApJ...653L.149K} and \citet{2013SoPh..284..405D} used the effect of photospheric albedo on the HXR spectrum to conclude that nonthermal electron distribution is close to isotropic, while inconsistent with a noticeable beaming. However, this conclusion pertains to the chromospheric target volume, rather than to the coronal source. \citet{Fl_2005n} employed microwave polarimetry to reveal a loss-cone anisotropy of trapped component of the nonthermal electrons, while \citet{Altyntsev_etal_2008, Altyntsev_etal_2016} and \citet{2014cosp...40E2068M} reported the beam-like anisotropy for some events. It is interesting that \citet{Altyntsev_etal_2008, Altyntsev_etal_2016} found the beam-like anisotropy in smaller loops in two events involving interaction between two different loops, which is in line with the finding discussed here.}

{Due to the beamed angular distribution most of the streaming nonthermal electrons} \blank{, so most of them}  immediately precipitate into the southern footpoint of the small loop and produce the HXR emission there. On the fly, they interact with the magnetic field of the loop, which is reasonably strong in the small loop, varying from $B\sim600$~G at the looptop up to $B\sim1200$~G at the footpoints, to produce the high-frequency \mw\ emission as observed. The total number of fast electrons, $N_{\rm r,tot}\approx1.35\times10^{34}$, needed to match the high-frequency part of the \mw\ spectrum at the peak time requires a roughly double acceleration rate as compared with that derived from the HXR thick-target model fit, $\sim1.2\times10^{35}$~electron/s.

The missing electrons, those not seen via HXR emission, must have escaped to the big loop and be trapped there. To confirm this quantitatively, we note that at the decay phase time frame 01:35:24.500~UT, which we analyzed in great detail to validate the model, the total number of the trapped fast electrons was found to be $N_{\rm r,tot}\approx5.7\times10^{34}$ to match the \mw\ spectrum. This implies that at the peak time of the gradual \mw\ light curves (01:35:05~UT), when the flux density at 3.75~GHz is twice bigger than at 01:35:24.500~UT, the number of the nonthermal electrons in the big loop must have been a factor of two larger, $N_{\rm r,tot}\approx1.2\times10^{35}$. This peak number of the fast electrons accumulated in the big loop is to be compared with the corresponding electron injection into the big loop. If we assume that the electron injection rate into the big loop is equivalent to the electron loss rate derived from the HXR thick-target spectral fit, the total number of electrons injected into the big loop would be $N_{\rm inj}\sim6\times10^{35}$ electrons over the impulsive phase of the flare; which, taken at the face value, is roughly five times larger than needed to supply the observed \mw\ emission from the big loop. Given that the number of the nonthermal electrons in the big loop is determined using a poorly defined low-energy spectral index and low-energy cut-off in the big loop, we conclude that the obtained electrons numbers are consistent with each other and so having a half of the accelerated electrons or slightly less to escape towards the big loop is sufficient to supply it with the required number of the fast electrons needed to match the low-frequency part of the \mw\ spectrum.

This picture is also quantitatively consistent with the light curves for various frequencies and energies. Indeed, the close correlation between the HXR (or 35~GHz) light curves and the time derivative of the 3.75~GHz light curve is consistent with the former being a proxy of the acceleration/injection time profile, while the latter being a proxy of the trapped electron component. Now, the delay in the SXR GOES light curves becomes transparent: the direct losses of the accelerated electrons immediately available for plasma thermal response (including heating and evaporation) are limited to only a roughly half of all electrons, which precipitate through the small loop. The other half of the accelerated electrons trapped in the big loop lives longer and continues to heat the plasma via \textit{in situ} Coulomb collisions in the loop and precipitation. This is why the GOES flux and the GOES-derived temperature continue to grow well after the impulsive phase of the flare is over. We emphasize that such a scenario is only possible if the accelerated electrons are roughly equally divided between these two loops as is the case in our model, which additionally confirms the fast electron numbers obtained above for these two loops from the independent \mw\ spectrum fits.

Having the model validated, we can now address a number of fundamental questions about magnetic reconnection, particle acceleration, and transport. Recall, that the coronal \mw\ source moves quickly with the apparent velocity of roughly 320~km/s passing about $35''$ over 80~s during the impulsive phase, which is reasonable to associate with a spread of the magnetic reconnection between the small and big loops. The process of magnetic reconnection will form new closed field lines (flux tubes), where the magnetic flux $\Phi$ is conserved along the field line \citep[see, e.g.,][]{Qiu_etal_2009} such as $\Phi\sim V_r B_r\tau L_r\sim V_{\rm lt}B_{\rm lt}\tau L_{\rm lt}$, where $V_r$, $B_r$, and $L_r$ are the velocity, magnetic field, and spatial scale at the reconnection site (site of interaction between the loops), $ V_{\rm lt}$, $B_{\rm lt}$, and $L_{\rm lt}$ are the same at the looptop, $\tau$ is the time of the reconnection process.
Given that we know the magnetic field at the loop top from the modeling, $B_{\rm big,lt}\approx 30$~G, the looptop source velocity, and displacement, we can estimate the reconnecting electric field as $E_r[{\rm V/cm}] \sim 10^{-3} B_{\rm lt}[{\rm G}]V_{\rm lt}[{\rm km/s}]\sim 10~{\rm V/cm}$. Since the magnetic field at the small loop is more than two orders of magnitude larger than $B_{\rm lt}\approx 30$~G, the expected displacement of the reconnection site along the small loop is within $3''$ and unobservable in agreement with the constant location of the southern \mw\ source. Perhaps, this spatial extent of the reconnection is responsible for the increase of the small loop width required to get a good \mw\ fit at the decay phase.

This process of magnetic reconnection results, directly or indirectly, in acceleration of a significant numbers of fast electrons to relativistic energies on a subsecond time scale. Interestingly, the \mw\ polarization data require that the accelerated electron distribution in the small loop is beamed along the magnetic field vector (i.e., from the northern source having the positive, north, magnetic polarity towards the southern source, having the negative, south, magnetic polarity). Recall, that the small loop has a negative $\alpha$, which implies that the electric field vector is directed oppositely to the magnetic field vector; thus, the fast electrons are beamed in the direction, where they are driven by the electric field; thus, the electric field is a likely cause of this electron beaming towards the southern footpoint.

The thermal electron number density in the small loop is poorly constrained; but given the fast electron number density is about $10^9$~cm$^{-3}$, the total electron number density is at least that big. On the contrary, we can get a good estimate of the thermal electron number density at the big loop using the \mw\ spectral shape at the late decay phase. Indeed, the spectral peak frequency stays constant late in the event, $\approx 1.5$~GHz, while the low-frequency spectral index is rather large, $\alpha_{\rm lf}\approx 3$, which are collectively indicative that the spectral peak is formed by Razin-effect $f_{\rm peak}\approx f_R$, rather than optical thickness effect, at the late decay phase. Given that the Razin frequency $f_R\approx 20 n_0/B$ and $B\approx 30$~G, we estimate the mean thermal number density at the big loop as $n_0 \sim (2-3)\times10^9$~cm$^{-3}$, which agrees well with the developed 3D model ($n_0 = 5\times10^9$~cm$^{-3}$ at the central field line of the loop and it decreases in the transverse direction over a gaussian law).

Note that the collisional loss time in such a tenuous plasma is longer that one minute for all electrons with energies higher than 100~keV responsible for \mw\ emission, while the observed decay time of the light curve at 3.75~GHz, for example, is about $\tau\sim30$~s. This unambiguously suggests that the high-energy fast electron loss from the big loop is mediated by the process of enhanced electron pitch-angle scattering by turbulence and their escape from the loop via the loss-cone. This is independently confirmed by lack of the electron spectral flattening, which must be present in case of collisionally-mediated electron transport.

In contrast, the low-energy electrons around the nominal low-energy cut-off of 10~keV are likely strongly affected by  in situ Coulomb losses. Indeed, the collisional loss time of 10~keV electron in a plasma with number density around $2\times10^9$~cm$^{-3}$ is about 3~s. The corresponding energy deposition to the coronal plasma is roughly $\dot{E}=\int\limits_{E_{\min}}^{E_{\max}}n_r(E)E/t_E dE \approx0.2$~erg~cm$^{-3}$~s$^{-1}$, where $n_r(E)$ is the nonthermal electron distribution over energy and $t_E$ is the collisional loss time \citep[cf. Sec.~4 in][]{Bastian_etal_2007}. Accordingly, the temperature increase over the interval $\tau$ is $\Delta T \sim \frac{\tau\dot{E}}{3n_0 k_B}$, where $k_B$ is the Boltzman constant; taking the observed duration of the main heating phase $\tau=30$~s and plugging in other relevant numbers, we obtain $\Delta T \sim7$~MK in agreement with GOES data. Thus, the energy-containing low-energy electrons deposit their energy directly to the coronal plasma, while the less numerous precipitating higher-energy electrons do not  deposit sufficient energy to the footpoints to drive efficient chromospheric evaporation.
This explains why we have a relatively strong \mw\ burst (which statistically corresponds to a M4-class event), but a rather weak C5 GOES flare.
We conclude, that we have obtained a fully consistent picture of this cold flare event.

\section{Conclusions}

In this study we identified a new ``cold'' solar flare whose properties and physical model are substantially different from the cold flares reported so far \citep{Bastian_etal_2007, Fl_etal_2011, 2013PASJ...65S...1M}. In contrast to the known cold flares, which consisted of one main loop, the described here 2002-03-10 cold flare is a vivid example of interaction between two loops. The first of them, a small one, is responsible for the impulsive flare component, while the bigger one is responsible for a more gradual nonthermal emission. Interestingly, the electrons accelerated in the event divided roughly evenly between these two loops, which made both loops comparably important in driving the thermal response in this event. For this reason the GOES flare was substantially delayed relative to the impulsive peak in apparent contradiction with the conventual Neupert effect. {Deviations from the nominal Neupert effect have widely been reported \citep[e.g.,][]{2002A&A...392..699V, 2003AdSpR..32.2459D, 2005ApJ...621..482V, 2008AdSpR..41..988S} and often interpreted as an evidence in favor of an additional source of plasma heating.} However, {no additional heating is needed to understand the heating delay in our event:} taking into account in situ coronal losses of the fast electron component trapped in the big loop, we obtained a scenario fully consistent with the plasma heating by the accelerated electrons---in a remarkable agreement with spirit of the Neupert effect. The developed model is in quantitative agreement with observations, including \mw\ imaging and polarization, and naturally identifies the cause of the suppressed chromospheric evaporation that is needed to interpret the unusually weak GOES response in this flare.

\acknowledgments

This work was supported in part by NSF grants  AGS-1250374 and AGS-1262772, NASA grant NNX14AC87G to New Jersey
Institute of Technology and RFBR  grants 15-02-01089, 15-02-03717, 15-02-03835,  15-02-08028, and 16-02-00749.
This study was supported
by the Program of basic research of the RAS Presidium No. 9. Authors
acknowledge the Marie Curie PIRSES-GA-2011-295272 RadioSun project.
We thank Dr. Gelu Nita for encouraging discussions.

%+++++++++++++++++++++++++ Figures+++++++++++++++++++++++++++++

%=========================End of Figures=========================================

\bibliographystyle{apj}
\bibliography{cold_flare_ref,fleishman,lkk_ref,neupert}

\appendix
\section{Thermal model for the gradual flare component}
\label{S_therm_model}

Let us consider a model in which the thermal emission plays a role at low frequencies, while the nonthermal gyrosynchrotron emission dominates the high frequencies although it can give some contribution at the low frequencies. There are two main mechanisms of the thermal emission at the \mw\ range---free-free and gyro emission (gyroresonance, GR, or gyro\-syn\-chrot\-ron, GS). Note that the opacity of the free-free emission \textit{decreases} as the plasma temperature \textit{increases}; thus, the plasma heating alone  results in a decrease of the \mw\ free-free emission; its increase requires simultaneous significant increase of the plasma density to the numbers inconsistent with the emission measure (\EM) estimate available from the SXR GOES data{; the peak value is $\EM\approx 2\times10^{48}$~cm$^{-3}$}. Therefore, if the observed emission is thermal it can only be the gyro emission.

The weak polarization at 1 \& 2~GHz tells us that the GR emission at these frequencies must be optically thick, while a significantly stronger polarization at 3.75 \& 5.7~GHz implies that only the dominant X-mode (LCP in our case) may remain thick, while the O-mode (RCP) is getting thin here. As long as the thermal emission remains optically thick, its flux in each of the eigen-modes (X and O) at a given frequency $f$ is firmly specified by a product of the plasma temperature $T$ and the source area $A$, such as

\begin{equation}
\label{Eq_Therm_Flux}
  F_{\rm LCP}\simeq F_{\rm RCP}\simeq 6~[{\rm sfu}] \left(\frac{f}{1~{\rm GHz}}\right)^{2}\left(\frac{T}{10^7~{\rm K}}\right)\left(\frac{A}{10^{20}~{\rm cm}^2}\right),
\end{equation}
and the total flux is equal to the sum of the two components, $F=F_{\rm LCP}+F_{\rm RCP}$.

Let us consider first the implications of the thermal model for the emission at 5.7~GHz, where we have imaging data needed to estimate the source sizes and area.
As has been estimated from Figure~\ref{figEIT}, left, the source sizes are $40''\times90''$, i.e., $A\simeq2\times10^{19}$~cm$^2$, while $F_{\rm LCP}\simeq  100~{\rm sfu}$; thus, Eq.~(\ref{Eq_Therm_Flux}) yields the plasma temperature around $T\sim25$~MK. This number looks somewhat excessive compared with the GOES-derived temperature, Figure~\ref{fig_LC}d,e; however, it can still be fine if the plasma is tenuous and the corresponding  \EM\ is small. Let us estimate the thermal plasma number density from the light curve cooling profile (the 5.7~GHz light curve, not shown in Figure~\ref{fig_LC}, is similar to that at 3.75~GHz). It is easy to estimate that the radiative cooling time is much longer than the observed decay time scale $\tau_{5.7{\rm GHz}}\sim30$~s; thus the cooling must be controlled by the thermal conduction, whose time scale is \citep[cf. Eq. (4.3.10) in][]{Aschw_2005}

\begin{equation}
\label{Eq_conduct_decay_time}
   \tau\simeq2.4\cdot10^3~{\rm [s]}
    \left(\frac{L}{10^{10}~{\rm cm}}\right)^{2}\left(\frac{n_e}{10^{10}~{\rm cm}^{-3}}\right)
   \left(\frac{10^7~{\rm K}}{T}\right)^{5/2},
\end{equation}
provided that the heat conduction has not yet reached the free-streaming limit.
Therefore, to obtain the right time scale of the conductive cooling from the observed loop with the length $L\sim10^{10}$~cm and temperature $T\sim25$~MK requires $n_e\sim10^9$cm$^{-3}$; being combined with the loop volume (assuming the sizes of $40''\times40''\times100''$ as observed) this density yields the emission measure $EM\sim10^{46}$~cm$^{-3}$, which is more than two orders of magnitude smaller than the GOES-derived background value, thus no GOES response is expected from this hot plasma. We conclude that the thermal model of the emission at 5.7~GHz does not contradict available observational constraints.

Having said that, we have yet to check if this conclusion holds at lower frequencies of which we concentrate now on a lower one, 1~GHz. %{Q to Larissa: can you check the RSTN data at 1.4 GHz and below?}.
The problem is that radio flux from a uniform thermal source scales as $\propto f^2$ with frequency, see Eq.~(\ref{Eq_Therm_Flux}), while we observe almost flat spectrum, which implies that the product $A\cdot T$ must scale roughly as $\propto f^{-2}$. We do not have spatially resolved measurements at 1~GHz, so
let us first consider the case of the same temperature $T\sim25$~MK but bigger area  $A\sim4\cdot10^{20}$~cm$^2$ needed to provide the observed flux density at 1~GHz.

\begin{figure}\centering
%C:\GS_Modeling\ThermalOVSA_Flares\150401_1\Uniform_Thermal_Spectrum_64.pro
%file='c:\MyProjects\ColdFlare10mar2002\thermal_spectrum_25MK.ps'
\includegraphics[width=0.6\columnwidth]{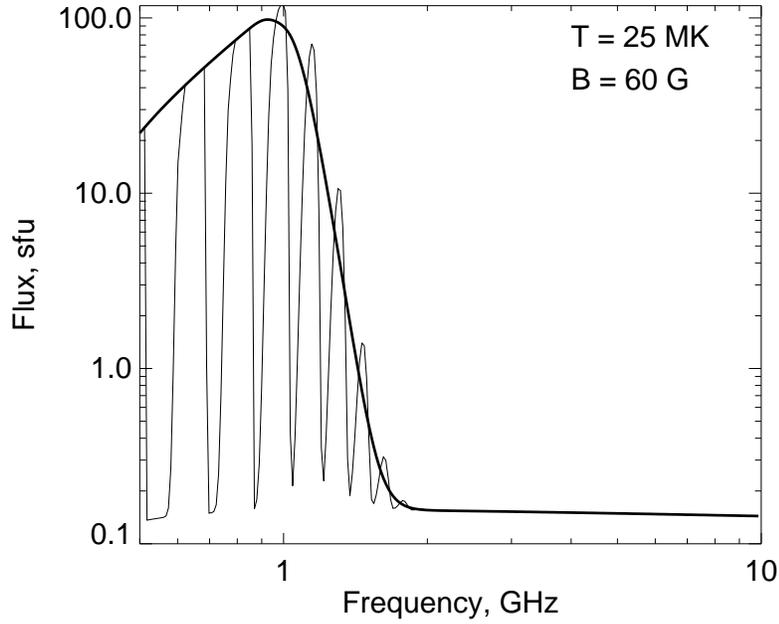}
\caption{\label{f_th_GS_spec_T25_B60} GS thermal spectrum plotted for a source with area $A=4\times10^{20}$~cm$^2$, depth $d=7\times10^8$~cm, $T=25$~MK, thermal density $n_0=10^9$~cm$^{-3}$, $B=60$~G, and the viewing angle $\theta=75^\circ$.  Thin curve shows exact spectrum from a uniform source containing the gyroharmonics, while the thick curve shows the corresponding averaged spectrum obtained using the continuous fast code \citep{Fl_Kuzn_2010}. Smaller magnetic field will result in lower spectral peak frequency.
}
\end{figure}

Given the implied increase of the source size at 1~GHz compared with that at 5.7~GHz, the conductive cooling time is consistent with the observed value $\tau_{1{\rm GHz}}\sim 150$~s for roughly the same number density $n_e\sim10^9$cm$^{-3}$. However, the enhanced source area also implies an enhanced volume, $V\sim3\cdot10^{30}$~cm$^3$, which yields the emission measure of $EM\sim3\cdot10^{48}$cm$^{-3}$. Such an \EM\ would easily be revealed by GOES data on top of a comparable background value, which is not observed and, thus, calls the thermal model into question. On top of that, %a 3D modeling with the extrapolated magnetic data cube
to get an optically thick thermal GR emission with the brightness temperature of 25~MK from a given line of sight at 1~GHz, the line of sight must cross a volume element with $T\sim25$~MK and the magnetic field equal to or exceeding 60~G, see Figure~\ref{f_th_GS_spec_T25_B60}. Figure~\ref{figEIT}, right, shows the projected area of all lines of sight satisfying the condition $|B|>50$~G from the magnetogram directly, from which we can directly compute the maximum possible area of such a thermal gyro source as $A(>50$~G$)\approx4.77\cdot10^{19}$~cm$^2$, which is by almost one order of magnitude insufficient to reproduce the observed radio flux at 1~GHz. Our tests with 3D modeling using the extrapolated magnetic data cube confirm that even if we fill the entire data cube with the hot plasma having $T\sim25$~MK, the mismatch between the modeled and observed flux at 1~GHz is more than a factor of two regardless of the selected combination of the input parameters.
Having higher temperature would imply accordingly bigger density to keep the conductive cooling time the same as observed. But this enhanced density yields an enhanced \EM\ in progressive disagreement with the GOES data.
We conclude that the thermal model is not supported by the data.

\clearpage

%% the file, uncomment the \plotone, \plottwo, and
%% If you need a layout that cannot be achieved with \plotone or
%% \plottwo, you can invoke the graphicx package directly with the
%% \includegraphics command or use \plotfiddle. For more information,
%% please see the tutorial on ``Using Electronic Art with AASTeX" in the
%% documentation section at the AASTeX Web site, http://aastex.aas.org/
%%
%% The examples below also include sample markup for submission of
%% supplemental electronic materials. As always, be sure to check
%% the instructions to authors for the journal you are submitting to
%% for specific submissions guidelines as they vary from

%% This example uses \plotone to include an EPS file scaled to
%% 80% of its natural size with \epsscale. Its caption
%% has been written to indicate that additional figure parts will be
%% available in the electronic journal.

%\begin{figure}
%\epsscale{.80}
%\plotone{f1.eps}
%\caption{Derived spectra for 3C138 \citep[see][]{heiles03}. Plots for all sources are available
%in the electronic edition of {\it The Astrophysical Journal}.\label{fig1}}
%\end{figure}

\clearpage

%% Here we use \plottwo to present two versions of the same figure,
%% one in black and white for print the other in RGB color
%% for online presentation. Note that the caption indicates
%% that a color version of the figure will be available online.
%%

%\begin{figure}
%\plottwo{fl_2013_04_21_16_02_Konus_Wind_G1.eps}{sp_Konus_RHESSI_6_8-1.eps }
%\caption{\label{fig1}}
%\end{figure}

%% This figure uses \includegraphics to scale and rotate the still frame

%\begin{figure}
%\includegraphics[angle=90,scale=.50]{f3.eps}
%\caption{Animation still frame taken from \citet{kim03}.
%This figure is also available as an mpeg
%animation in the electronic edition of the
%{\it Astrophysical Journal}.}
%\end{figure}

\end{document}